\documentclass[12pt]{article}
\usepackage{epsfig,pstricks}
\usepackage{amsmath}
\usepackage{hhline}
\usepackage{amssymb}
\usepackage{times}
\usepackage{cite}

\newlength{\dinwidth}
\newlength{\dinmargin}
\newcommand{\qs}{\ensuremath{Q^2}}
\newcommand{\gevsq}{\ensuremath{\mathrm{Ge\hspace*{-1pt}V}^2}}
\newcommand{\etj}{\ensuremath{E_T^{\,jet}}}
\newcommand{\etaj}{\ensuremath{\eta^{\,jet}}}
\newcommand{\setj}{\ensuremath{{\rm d}\sigma / {\rm d}E_T^{\,jet}}}
\newcommand{\setaj}{\ensuremath{{\rm d}\sigma / {\rm d}\eta^{\,jet}}}
\newcommand{\gev}{\ensuremath{\mathrm{Ge\hspace*{-1pt}V}}}
\newcommand{\mev}{\ensuremath{\mathrm{Me\hspace*{-1pt}V}}}
\newcommand{\pb}{\ensuremath{\mathrm{pb}}}
\newcommand{\nb}{\ensuremath{\mathrm{nb}}}

\newcommand{\pbi}{\ensuremath{\mathrm{pb}^{-1}}}

\newcommand{\m}{\ensuremath{\mathrm{m}}}

\newcommand{\alp}{\ensuremath{\alpha_S}}

\newcommand{\ET}{\mbox{$E_T$}}
\newcommand{\LQ}{\ensuremath{\Lambda_{\mathrm{QCD}}}}
\newcommand{\pThm}{\ensuremath{\hat{p}_T^{\mbox{\scriptsize{~min}}}}}

\newcommand{\Wgp}{\ensuremath{W_{\gamma p}}}

\newcommand{\relbarx}{\ldots}

\def\costh3{\cos\theta_3}

\newcounter{fnun}
\newcounter{fndeux}
 
\begin{document}

%===============================title page=============================

\pagestyle{empty}
\begin{titlepage}
\begin{flushleft}
{\tt DESY 02-225    \hfill    ISSN 0418-9833} \\
{\tt February 2003}                  \\
\end{flushleft}
\vspace*{1cm}

\vspace*{1cm}
\begin{center}
  \Large
  {\bf Measurement of inclusive jet cross sections \\  
    in photoproduction at HERA}

\vspace*{2cm}
    {\Large H1 Collaboration} 
\end{center}

\vspace*{2cm}
\begin{abstract}

\noindent
Inclusive jet cross sections are measured in photoproduction at HERA using the H1 detector. 
The data sample of $e^+ p \rightarrow e^+ + jet + X$ events in the kinematic range of photon virtualities
$Q^2 \leq 1~\gevsq$ and photon-proton centre-of-mass energies $95 \leq \Wgp \leq 285~\gev$ represents an integrated luminosity of $24.1~\pbi$.
Jets are defined using the inclusive $k_{\bot}$ algorithm.
Single- and multi-differential cross sections are measured as functions of jet transverse
energy \etj ~and pseudorapidity \etaj ~in the domain $5 \leq \etj
\leq 75~\gev$ and $-1 \leq \etaj \leq 2.5$.
The cross sections are found to be in good agreement with next-to-leading order
perturbative QCD calculations corrected for fragmentation and underlying event
effects.
The cross section differential in \etj, which varies by six orders of magnitude over the measured range,
is compared with similar distributions from 
$p \bar{p}$ colliders at equal and higher energies.   

\end{abstract}
\vspace*{1.5cm}
 
\begin{center}
Submitted to Eur. Phys. J. C
\end{center}
 
\cleardoublepage
\end{titlepage}
 
%
 
%          COPY THE AUTHOR AND INSTITUTE LISTS
 
%       AT THE TIME OF THE T0-TALK INTO YOUR AREA
 
%
 
% from /h1/iww/ipublications/h1auts.tex

%
%
\begin{flushleft}
  %-- H1AUTS Author list by names 
%-- Status: Fri Jun  7 18:01:50 MET DST 2002  Number of authors = 321 
% !!! This version includes corrections up to 5/12/02  Joachim Meyer !!!!!

C.~Adloff$^{33}$,              %WUPP-LEFT      07/01           Adloff              
V.~Andreev$^{24}$,             %LPI -PD        8/88            Andreev             
B.~Andrieu$^{28}$,             %ECPL-LEFT      09/01           Andrieu             
T.~Anthonis$^{4}$,             %ANTW-ST        11/99           Anthonis            
A.~Astvatsatourov$^{35}$,      %ZEUT-ST        02/98           Astvatsatourov      
A.~Babaev$^{23}$,              %ITEP-PD        8/88            Babaev              
J.~B\"ahr$^{35}$,              %ZEUT-PD        8/88            Baehr               
P.~Baranov$^{24}$,             %LPI -PD        8/88            Baranovp            
E.~Barrelet$^{28}$,            %PARI-PD        11/99           Barrelet            
W.~Bartel$^{10}$
% ...

%-- H1 Institutes 
\bigskip{\it
% ...
$ ^{4}$ Inter-University Institute for High Energies ULB-VUB, Brussels;
          Universiteit Antwerpen (UIA), Antwerpen; Belgium$^{ c}$ \\
% ...
$ ^{10}$ DESY, Hamburg, Germany \\
% ...
$ ^{23}$ Institute for Theoretical and Experimental Physics,
          Moscow, Russia$^{ l}$ \\
 $ ^{24}$ Lebedev Physical Institute, Moscow, Russia$^{ e}$ \\
% ...
$ ^{28}$ LPNHE, Universit\'{e}s Paris VI and VII, IN2P3-CNRS,
          Paris, France \\
% ...
$ ^{33}$ Fachbereich Physik, Bergische Universit\"at Gesamthochschule
          Wuppertal, Wuppertal, Germany \\
% ...
$ ^{35}$ DESY, Zeuthen, Germany \\
% ...
$ ^c$ Supported by FNRS-FWO-Vlaanderen, IISN-IIKW and IWT \\
% ...
$ ^e$ Supported by the Deutsche Forschungsgemeinschaft \\
% ...
$ ^l$ Partially Supported by Russian Foundation
      for Basic Research, grant    no. 00-15-96584 \\
}

\end{flushleft}

\newpage
\setcounter{page}{1}
\pagestyle{plain}

\section{Introduction}

\begin{figure}[t]
\begin{center}
\vspace*{-0.3cm}
 \epsfig{file=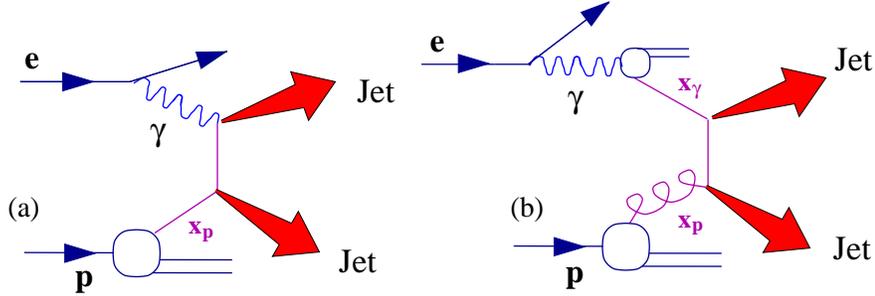,width=0.8\textwidth} 
\end{center}
\pspicture(-2.3,0.)(9,2) 
\rput[r](0,4){{\small (a)}}
\rput[r](6.7,4){{\small (b)}}
\endpspicture
\vspace*{-2.5cm}
\caption{\label{diagrams}Example LO QCD diagrams for inclusive jet
photoproduction in direct (a) and resolved (b) photon interactions.}
\end{figure}

At HERA, the interaction of protons with quasi-real photons emitted from the electron\footnote{
In the data taking periods used for this analysis, HERA was operated with a positron beam. 
However, the generic name ``electron'' will be used here 
to mean both electron and positron.
}
beam can result in the production of jets~\cite{Ahmed:1992xj,Derrick:1992zd},
for which two types of process are responsible in leading order (LO) quantum
chromodynamics (QCD).
The photon may interact as a pointlike particle with a parton carrying a
fraction $x_p$ of the proton momentum,
in so-called direct processes (Fig.~\ref{diagrams}(a)).
Alternatively, in resolved processes (Fig.~\ref{diagrams}(b)), the photon may
develop a hadronic structure so that a parton carrying a fraction $x_\gamma$ of
the photon momentum
interacts with a parton in the proton.
Due to confinement, the partons emerging from the interaction fragment into
jets of colourless particles. 
The hadronic final state also includes
the proton remnant and, in the case of resolved processes, the photon remnant and additional particles resulting from possible remnant-remnant interactions (the underlying event).

\noindent
The main purpose of this paper is to present inclusive jet 
cross sections measured using the H1 detector and to compare them
with fixed order parton level  QCD
predictions. After correcting the data and calculations to the hadron level, these comparisons offer a
means of testing the validity of perturbative QCD predictions, including the
description of the partonic structure of the photon and the proton in terms of
parton distribution functions (PDFs). The QCD-inspired modelling of
non-perturbative effects in hard hadronic photoproduction can be tested as
well.
The most accurate theoretical predictions have been calculated up to
next-to-leading order (NLO) in perturbative QCD. In order to compare these
predictions with jet cross section measurements, the jet definition must meet
certain requirements, such as infrared and collinear safety and
minimal sensitivity to 
fragmentation and underlying event effects.
The $k_{\bot}$ algorithm, originally proposed in \cite{Catani:1991hj}, satisfies these requirements.

\noindent
Inclusive jet measurements at hadron colliders~\cite{Albajar:1988tt,Arnison:1986vk,Abbott:2001ce,Abbott:1998ya,ppbarjet}
and at HERA~\cite{Abt:1993mx,Derrick:1994aq,Aid:1995ma,Breitweg:1998ur,unknown:2002ru} have
often been important in the development of the understanding of QCD.
In this paper, the first H1 measurement of inclusive jet cross sections in
photoproduction using the inclusive $k_{\bot}$ algorithm~\cite{Ellis:1993tq} is 
presented. 
Compared with the last H1 inclusive jet
measurement~\cite{Aid:1995ma},  the integrated luminosity has been increased by
a factor of $80$ and the jet transverse energy range has been extended, now
reaching from $5$ to $75~\gev$. 
In order to allow cross-checks with previous HERA measurements and comparisons with data from other colliders, the measurements have also been performed using a cone algorithm.

\noindent
The paper is organized as follows.
In section~\ref{pheno}, the motivations for this measurement are detailed and the phenomenology of inclusive jet photoproduction is presented.
A brief description of the H1 detector and details of the analysis procedure are given in section~\ref{expe}.
The measurements of single- and multi-differential inclusive jet cross sections
as functions of jet transverse energy \etj ~and pseudorapidity\footnote{
Pseudorapidity is defined as $\eta \equiv -\ln(\tan~\theta / 2)$, where $\theta$ is the polar angle,
in the coordinate system centered at the nominal interaction point, with respect to the positive $z$ axis along the outgoing proton beam direction.
The outgoing 
proton (photon) direction is also called forward (backward) and corresponds to $\eta > 0$ ($\eta < 0$).
}\setcounter{fnun}{\thefootnote}
\etaj ~in the laboratory frame are presented in section~\ref{result}.
The results are compared with LO and NLO QCD calculations and
with inclusive jet measurements at 
$p \bar{p}$ colliders. 
The final section provides a summary of the results.

\section{Inclusive jet photoproduction}
\label{pheno}

Inclusive jet cross sections are obtained by counting the number of jets found by a jet algorithm in a given kinematic range.
The inclusive $k_{\bot}$ algorithm~\cite{Ellis:1993tq} is a modified version of the exclusive $k_{\bot}$ algorithm~\cite{Catani:1993hr} where all hadronic final state particles are clustered iteratively\footnote{For more details, see e.g.~\cite{Adloff:1998ni}.} into jets according to their separations in $(\eta,\phi)$ space\footnote{$\phi$ is the azimuthal angle in the transverse plane.}.
Here, the algorithm is applied in the laboratory frame.
The separation parameter between jets in $(\eta,\phi)$ space is set to $D = 1$, as in~\cite{Adloff:2002au}.
An \ET\ weighted recombination scheme~\cite{snowmass}, in which the reconstructed jets are massless, is used to maintain invariance under longitudinal boosts.
To allow comparisons of the results of this study with previous measurements in photoproduction and with other collider data, the complete analysis has also been performed using a cone algorithm~\cite{jetalgcone} with a cone radius $R = 1$. 
The cone algorithm has been shown to give larger hadronisation corrections than the inclusive $k_{\bot}$ algorithm (section~\ref{hadronisation}), as is the case for jets in deep inelastic scattering (DIS)~\cite{Adloff:2001tq}.

\subsection{Motivation}

In this paper, inclusive jet cross sections are measured 
over a very large \etj ~range.
In the high \etj ~region, 
the high transverse momentum of the outgoing parton provides a hard scale which allows reliable cross
section calculations to be made in perturbative QCD.
It also ensures a reduced influence of less-well understood soft processes (fragmentation and underlying event).
Jets at high \etj ~thus provide the most direct insight into photoproduction at the parton level.

\noindent
In the region of low \etj, the NLO and higher order terms as well as corrections from the parton to
the hadron level become more important, since the strong coupling \alp ~increases with decreasing 
scale. In the absence
of a fundamental understanding of 
non-perturbative processes, the comparisons between data and theory necessarily involve
phenomenological models. Matching the theoretical predictions with the experimental measurements at low
\etj ~thus represents a further important test of QCD-inspired phenomenology in jet
photoproduction.

\noindent
Jet photoproduction cross sections are directly sensitive 
to the gluon as well as the quark content of the photon and the proton.
The proton PDFs are precisely determined~\cite{Lai:1999wy,Martin:2000ww,Gluck:1994uf} from
structure function measurements ~\cite{Breitweg:2000yn,Adloff:2000qk}.
An exception is the gluon distribution at high
$x_p$~\cite{Pumplin:2002vw}. There, jet photoproduction
measurements~\cite{Chekanov:2001bw,Adloff:2002au} can provide complementary
information. For the photon, the quark density at medium and high $x_{\gamma}$
is constrained~\cite{Gluck:1992jc,Gluck:1992ee,Aurenche:1994in} by $F_2^{\gamma}$ measurements at $e^+e^-$ colliders~\cite{Krawczyk:2001mf}, albeit
with larger uncertainties than in the proton case.  Since boson-gluon fusion is
a higher order process compared with photon-quark scattering at $e^+ e^-$
colliders, the gluon density in the photon is even less well constrained.
Furthermore, the photon PDFs do not obey strict momentum sum rules, so that
there is no indirect constraint on the gluon density in the photon. In
photoproduction at HERA, higher scales can be reached than 
at LEP because of the higher centre-of-mass (cms) energy and higher $\gamma p$ compared with
$\gamma \gamma$ 
luminosity. 
Jet photoproduction
cross section measurements~\cite{unknown:2002ru,Chekanov:2001bw,Adloff:2002au,Adloff:2000bs} 
thus access a largely unexplored domain of photon structure.

\noindent
The measurement presented here probes a wide range of $\etj$ and $\etaj$,
quite  similar to the range explored in a recent dijet cross section
measurement~\cite{Adloff:2002au}. Although dijet events offer better
constraints on the hard scattering kinematics, inclusive jet measurements offer
the experimental advantages of greater statistics and increased kinematic range
and the theoretical advantage of naturally avoiding infrared-sensitive regions
of phase space~\cite{Frixione:1997ks}.

\subsection{QCD calculations}
\label{phenoNLO}

By considering the electron as a source of quasi-real photons of
virtuality\footnote{The photon virtuality is $Q^2 \equiv -(k - k')^2$, where
$k$ ($k'$) is the 4-vector of the incoming (outgoing) electron.}
$\qs$ and energy $E_{\gamma}$ and using the QCD factorisation theorem and 
a factorisable jet algorithm, the electron-proton cross section for producing $N$ jets ($N \geq 2$) in direct photoproduction can be expressed as:
\begin{equation}
\label{sigmaep_dir}
\sigma^{ep\rightarrow e + N~{\rm jets} + X}_{\rm dir.}=\int_{\Omega} {\rm d}\Omega\ f_{\gamma/e}(y,\qs)\ \sum_{i}\ f_{i/p}(x_p,\mu^2_p) \ {\rm d}\hat{\sigma}(\gamma \,i \rightarrow N~{\rm jets})~.
\end{equation}
Here, $y=E_{\gamma}/E_e$ is the fraction of the electron energy $E_e$ carried
by the photon, 
$f_{\gamma /e}(y,\qs)$ is the photon flux associated with the
electron~\cite{wwa}, 
$f_{i/p}(x_p,\mu^2_p)$ is the proton PDF of parton~$i$ evaluated at the factorisation scale $\mu_p$ and $\hat{\sigma}(\gamma \,i \rightarrow N~{\rm jets})$ is the parton-level cross section for the direct subprocess $\gamma \, i \rightarrow  N~{\rm jets}$.
The cross section $\hat{\sigma}$ is proportional to $\alpha_{em}(\mu^2_R) \alp^{N - 1}(\mu^2_R)$ at lowest order and can be expanded in powers of \alp ~multiplied by perturbatively calculable coefficient functions, both of which depend on the renormalisation scale $\mu_R$.
The kinematic domain over which the cross section  is
integrated is denoted $\Omega$.

\noindent 
Similarly, the cross section for resolved photoproduction can be written:
\begin{equation}
\label{sigmaep_res}
\sigma^{ep\rightarrow e + N~{\rm jets} + X}_{\rm res.}=\int_{\Omega} {\rm d}\Omega\ f_{\gamma/e}(y,\qs)\ \sum_{i \, j}\ f_{i/p}(x_p,\mu^2_p)\ f_{j/\gamma}(x_{\gamma},\mu^2_{\gamma}) \ {\rm d}\hat{\sigma}(i\,j \rightarrow N~{\rm jets})~.
\end{equation}
Compared with Eq.~\ref{sigmaep_dir} for direct processes, the resolved cross
section in Eq.~\ref{sigmaep_res} includes in addition the photon PDF of
parton~$j$, $f_{j/\gamma}(x_{\gamma},\mu^2_{\gamma})$, evaluated at the
factorisation scale $\mu_{\gamma}$. 
Due to the splitting $\gamma \rightarrow q \bar{q}$, the QCD evolution
equations of the resolved photon PDFs \cite{Witten:ju} differ from those for 
the proton and lead to large quark densities at high $x_\gamma$, which 
increase with $\mu_\gamma$. 
The cross section $\hat{\sigma}(i \, j
\rightarrow N~{\rm jets})$ describes the resolved subprocess $i \, j
\rightarrow N~{\rm jets}$ and is proportional to $\alp^{N}(\mu^2_R)$ at lowest order.

\pagebreak
\noindent 
The distinction between the direct (Eq.~\ref{sigmaep_dir}) and the
resolved (Eq.~\ref{sigmaep_res}) components is only unambiguous at LO,
whereas
beyond LO their relative contributions 
depend on the factorisation scale $\mu_{\gamma}$.
The inclusive cross section for the production of a jet in a given kinematic
range
$\sigma^{ep\rightarrow e + {\rm jet} + X}$
is obtained by summing all calculated contributions of
Eqs.~\ref{sigmaep_dir} and~\ref{sigmaep_res} over $N$, 
weighting by the corresponding number of jets inside this kinematic range.

\noindent 
The partonic cross sections $\hat{\sigma}$ in Eqs.~\ref{sigmaep_dir} and
\ref{sigmaep_res} have been calculated at LO and NLO in QCD by several
the\-oretical
groups~\cite{Frixione:1997ks,Klasen:1997it,Harris:1997hz,Aurenche:2000nc}.
These calculations 
differ mainly in the treatment of infrared and collinear singu\-larities.
In this paper, measurements are compared with the LO and NLO
calculations of~\cite{Frixione:1997ks}, based on the subtraction
method as implemented in a Monte Carlo program~\cite{Frixione:1997np}. This program generates weighted parton kinematic 
configurations used as input to the inclusive $k_{\bot}$
algorithm. These calculations differ from those obtained with
the phase space slicing method~\cite{Aurenche:2000nc} by less than $2\,\%$~\cite{Ferron:2001}.

\noindent
The CTEQ5M~\cite{Lai:1999wy} parameterisation of the proton PDFs was used for
the calculations. To test the dependence of the NLO cross sections on the
choice of proton PDFs, MRST99~\cite{Martin:2000ww} and CTEQ5HJ~\cite{Lai:1999wy}
were also used, where the latter has an enhanced gluon distribution at high
$x_p$. The renormalisation group equation to $2$-loop
accuracy was used for \alp ~and the value of \LQ ~was chosen to match that used
in the evolution of the proton PDFs (for five quark flavours, $\alp (M_Z) =
0.118$ for CTEQ5M and CTEQ5HJ, $\alp (M_Z) = 0.1175$ for MRST99).
GRV-HO~\cite{Gluck:1992ee} was chosen as the standard parameterisation of the
photon PDFs. The AFG-HO~\cite{Aurenche:1994in} parameterisation was also used to study the dependence of the
results on the choice of photon PDFs. The  renormalisation and factorisation
scales were defined as the sum of the transverse energies of the outgoing
partons divided by $2$.
These scales were varied by factors $0.5$ and $2$ in order to estimate the uncertainty corresponding to the missing higher-order terms.

\subsection{Monte Carlo models}
\label{phenoMC}

Simulated event samples have been used to correct the data for detector effects (section~\ref{crossdet}) 
and to estimate hadronisation effects for the QCD calculations
(section~\ref{hadronisation}).
Direct and resolved photoproduction events were simulated using the 
PYTHIA~\cite{Sjostrand:1994yb}, HERWIG~\cite{Marchesini:1992ch}, and
PHOJET~\cite{phojet} Monte Carlo generators. The generated events were passed
through a GEANT~\cite{Brun:1987ma} simulation of the H1 detector and
the same reconstruction chain as the data.

\noindent
All programs generate partonic interactions using the Born level QCD
hard scattering matrix elements, regulated by a minimum cut-off \pThm ~on the
common transverse momentum of the two outgoing partons. For PYTHIA
and PHOJET (HERWIG), the strong coupling constant \alp ~was calculated by
solving the $1$ ($2$) loop  renormalisation group equation
using $\LQ =200~\mev$ for $4$ ($5$) quark flavours.
GRV-LO parameterisations of the proton~\cite{Gluck:1994uf} and the photon~\cite{Gluck:1992jc} PDFs were used.
Higher order QCD radiation effects are simulated through initial and final state parton showers in the leading log approximation.
The fragmentation process is performed using the Lund string model~\cite{lund} as implemented in JETSET~\cite{jetset} in the case of PYTHIA and PHOJET and using a cluster model~\cite{cluster} in the case of HERWIG.

\noindent
For resolved photon interactions, besides the primary parton-parton scattering,
additional interactions are generated in order to simulate the effect of the
underlying event. \nopagebreak Within PYTHIA, these are calculated as LO QCD processes
between partons \pagebreak from the remnants of the proton and the photon. The resulting
additional final state partons are required to have transverse momenta above 
$1.2~\gev$, a value which gives an
optimal description of the transverse energy flow outside
jets for the specific photon PDFs used~\cite{Aid:1995ma}. Soft particles
accompanying the hard subprocess are produced in HERWIG using a soft underlying
event (SUE) mechanism which is based on parameterisations of experimental results
on soft hadron-hadron collisions. The fraction of resolved interactions which
are generated with an additional SUE can be varied within HERWIG and has been
adjusted to $35\,\%$ to match the observed level of soft activity between jets.
PHOJET, which is based on the two-component dual parton model~\cite{dpm}
incorporates detailed simulations of multiple soft and hard parton
interactions on the basis of a unitarisation scheme. Due to this scheme, small
variations of the lower momentum cut-off for hard parton interactions, set here
to $\pThm =3~\gev$, do not have an influence on the results of this generator.

\subsection{Hadronisation corrections}
\label{hadronisation}

Since the QCD calculations refer to jets of partons, whereas the measurements refer to jets of hadrons, the predicted cross sections have been corrected to the hadron level using LO Monte Carlo programs.
The hadronisation correction factors, $(1+\delta_{hadr.})$, are defined as the ratio of the cross sections obtained with jets reconstructed from hadrons after the full event simulation (including parton showers, fragmentation and underlying event effects) to that from partons after parton showers but before fragmentation and underlying event simulation. 
These corrections were calculated by taking the results from two different Monte
Carlo models chosen as described in section~\ref{crossdet}.
The uncertainty on these corrections was taken as the quadratic sum of the statistical error and the systematic error given by half the difference between the results obtained from the two Monte Carlo models. 
Using the $k_{\bot}$ algorithm, the corrections were found to be approximately
$30\,\%$ for $\etj < 10~\gev$ falling to values typically below $12\,\%$ for $\etj >
20~\gev$.
With the cone algorithm, the corrections are around $40\,\%$ for $\etj < 15~\gev$ and $20\,\%$ for $\etj > 15~\gev$.
The difference between the results obtained with the two Monte Carlo models is
typically very small and at most $10\,\%$.

\noindent The effects of the underlying event and of the fragmentation were also studied separately.
The corresponding correction factors, 
$(1+\delta_{u.e.})$ and $(1+\delta_{frag.})$, were determined in the same way as the overall corrections factors $(1+\delta_{hadr.})$.
Here, $(1+\delta_{u.e.})$ is defined as the ratio of the cross sections obtained
with jets reconstructed from hadrons with simulation of the underlying event to
that from hadrons without simulation of the underlying event, whilst
$(1+\delta_{frag.})$ is defined as the ratio of the cross sections obtained with jets reconstructed from hadrons to that from partons after parton showers, both without simulation of the underlying event.
By definition, $(1+\delta_{hadr.})=(1+\delta_{u.e.})\cdot(1+\delta_{frag.})$.

\noindent
Low momentum hadrons from the underlying event lead to a systematic increase of \etj ~and 
thus of the hadron level cross section at fixed \etj.
The $\delta_{u.e.}$ correction is always positive and increases as \etj
~decreases or \etaj ~increases.
Using the inclusive $k_{\bot}$ algorithm, for $5 \leq \etj < 12 ~\gev$,
$\delta_{u.e.}$ varies between $\sim 30 \, \%$ at $\etaj =-0.75$ and $100 \, \%$ at
$\etaj = 1.25$.
For $\etj > 20~\gev$,
$\delta_{u.e.}$ is always below $10 \, \%$.
The effect of the underlying event is partially compensated by fragmentation,
which has a tendency to lower the cross section. 
In general, $\delta_{frag.}$ is negative and becomes more important as \etj ~decreases but also as \etaj ~decreases, in contrast to  $\delta_{u.e.}$. 
The $\delta_{frag.}$ correction is around $ -
30 \, \%$ for $5 \leq \etj < 12 ~\gev$ and around $ - 5 \, \%$ for $\etj \geq
20~\gev$.

\section{Experimental technique}
\label{expe}

\subsection{H1 detector}

A detailed description of the H1 detector can be found elsewhere~\cite{deth1}.
Here only the components relevant for this measurement are briefly described.

\noindent 
The $ep$ luminosity is determined with a precision of $1.5\,\%$ by comparing
the event rate in the photon detector, located at $z= -103~\m$, with the cross
section for the QED bremsstrahlung process $ep \rightarrow ep \, \gamma$.
The scattered electron may be detected in the electron tagger (ETag), located at $z= -33~\m$.
Both detectors are TlCl/TlBr crystal \v{C}erenkov calorimeters with an energy resolution of $22 \, \% / \sqrt{E/\gev}$.

\noindent The central tracker (CT), which covers the range $ | \eta | \leq 1.5$
is used to measure the trajectories of charged particles and to reconstruct the interaction vertex.
It consists of inner and outer cylindrical jet chambers for precise position measurement in the transverse plane, $z$-drift chambers for precise $z$ measurement and proportional chambers for triggering purposes.
The transverse momentum of charged particles is reconstructed from the
curvature of tracks in the homogeneous magnetic field of $1.15$ Tesla along the
beam direction, with a resolution 
$\sigma (p_T)/p_T \approx 0.6 \, \% \, p_T/ \gev $.

\noindent
The finely segmented Liquid Argon (LAr) calorimeter~\cite{Andrieu:1993kh}
surrounds the tracking system and covers the range $ -1.5 \leq \eta  \leq 3.4$
with full azimuthal acceptance. It consists of an
electromagnetic section with lead absorbers and a hadronic section with
steel absorbers. The total depth of the LAr calorimeter ranges from $4.5$ to
$8$ hadronic interaction lengths. The energy resolution determined in test beam
measurements is 
$\sigma(E)/E \approx 50\,\% / \sqrt{E/\gev} \oplus 2\,\%$ for charged pions.
For jets with \etj ~above $20~\gev$, 
the jet energy calibration agrees at the $2\,\%$ level with the Monte Carlo simulation, as determined by the \ET\ balance 
in neutral current (NC) DIS 
and jet photoproduction events.
At lower \etj, the absolute hadronic energy scale is known  to $4\,\%$.
The absolute resolution in \etaj ~is approximately $0.05$ at $\etj = 5~\gev$ and better than $0.02$ for $\etj > 20~\gev$.

\noindent The 
region $ -4.0 \leq \eta  \leq -1.4$
is covered
by the SPACAL lead/scintillating-fibre calorimeter~\cite{Appuhn:1996na}.
Its absolute hadronic energy scale is known to $7\,\%$.

\subsection{Event selection and reconstruction}
\label{selec}

The data used in this paper were collected in $1996$ and $1997$, 
when electrons of energy $E_e = 27.5~\gev$ collided with
protons of energy $E_p = 820~\gev$, resulting in an $ep$ cms energy of
$300~\gev$.
For measurements in the region $\etj \geq 21 ~\gev$ (``high'' \etj), the full data sample representing an integrated luminosity of  $24.1~\pbi$ was used.
In addition to some activity in the central region, as seen by the CT, the trigger conditions required high transverse energy deposits in the LAr calorimeter (jet triggers).
In the region $ 5 \leq \etj  < 21 ~\gev$ (``low'' \etj), where jet triggers
suffer from proton beam-induced background, a trigger based on scattered
electron signals in the ETag was used instead.
This trigger was operated during a `minimum bias' data taking period corresponding to an integrated luminosity of $0.47 ~\pbi$.
The events from this subsample were required to have the scattered electron
detected in the fiducial volume of the ETag, with an energy in the
range $9.6 \leq E_{e}' \leq 19.3 ~\gev$.
The ETag geometrical acceptance, which is corrected for on an event-by-event
basis, is always greater than $10\,\%$ in this range.
The detection of the scattered electron ensures an improved measurement of $y$
and hence of the photon-proton cms energy $\Wgp = \sqrt{4 y E_e E_p}$,
but reduces the available number of events by a factor of approximately ten,
due to the restricted $y$ range and to the limited acceptance of the ETag.
The ETag events were also required to have no energy deposited in the photon arm of the luminosity system.
This condition suppresses background from high rate Bethe-Heitler events in random coincidence with proton beam-induced background events which give activity in the interaction region.
It also reduces QED radiative corrections.

\noindent 
An interaction vertex, reconstructed from tracks in the CT and located within
$30 \ {\rm cm}$ of the nominal $z$ position of the interaction point, was demanded.
Energy deposits in the calorimeters and tracks in the CT were combined, in a
manner that avoids double counting, 
in order to optimize the reconstruction of the hadronic final
state~\cite{Adloff:1997mi}, from which 
$\Wgp$
was derived~\cite{jb} for the \mbox{``high''}~\etj analysis.
The inclusive jet sample was then defined by keeping all events for which at least one jet was reconstructed with the inclusive $k_{\bot}$ algorithm in the kinematic domain:
\begin{equation}
\label{jetrange2}
-1 \leq \eta^{jet} \leq 2.5 ~;~ \etj \geq 21~\gev~~(\mbox{``high''}~\etj)~;
\end{equation} 
\begin{equation}
\label{jetrange1}
-1 \leq \eta^{jet} \leq 2.5 ~;~ \etj \geq 5~\gev~~(\mbox{``low''}~\etj).
\end{equation} 
The \etaj ~range was chosen to ensure that the jets were well contained in
the LAr calorimeter.
For ``high'' \etj ~events, the kinematic region was restricted to
\begin{equation}
\label{highETrange}
\qs \leq 1~\gevsq ~;~ 95 \leq \Wgp \leq 285~\gev. 
\end{equation} 
The \qs ~range is implied by the absence of the scattered electron in the main H1 detector.
For ``low'' \etj ~events, the tagging of the electron
already restricted the kinematic region to
\begin{equation}
\label{lowETrange}
\qs \leq {10}^{-2}~\gevsq ~;~   164 \leq \Wgp \leq 242~\gev. 
\end{equation} 
A number of requirements were made in order to suppress the non-$ep$ background.
For ``high'' \etj, the vertex condition was sufficient to reduce the contamination from beam-induced background to a negligible level.
Background originating from cosmic showers and beam halo muons was rejected using a set of topological muon finders\cite{Adloff:2000gv}.
In addition, the total missing transverse momentum ${P\hspace{-0.1in}/}_T$ was required to be small compared with 
the total transverse energy $E_T$ by applying the cut ${P\hspace{-0.1in}/}_T/\sqrt{E_T}\leq 2.5{~\gev}^{\frac {1}{2}}$.
The overall non-$ep$ background contamination in the ``high'' \etj ~sample was then estimated to be below $1\,\%$.
For ``low'' \etj ~events, since \Wgp ~can be measured in two independent ways
(using either the energy of the scattered electron or the hadronic final state), consistency between the results of the two methods was required.
By fitting the distribution of the $z$ position of the vertex with the sum of a
Gaussian and a constant, the ``low'' \etj ~sample was estimated to have a
non-$ep$ background contamination of about $2\,\%$.

\noindent
Further cuts were applied to reduce backgrounds from other $ep$ collision processes.
For the ``high'' \etj ~jet sample, the only significant one
is the NC DIS process\footnote{
The charged current DIS background was already completely suppressed by the ${P\hspace{-0.085in}/}_T$ cut.
}, where either the scattered electron or the current jet enters the inclusive jet selection.
Events with a scattered electron candidate found using electron identification algorithms~\cite{Adloff:2000qk} were thus rejected.
The remaining $ep$ background in the ``high'' \etj ~sample was estimated to be
below $1\,\%$ from a study using a simulated sample
of NC DIS events from the DJANGO~\cite{Django} Monte Carlo generator. 
In the ``low'' \etj ~sample it is completely negligible.

\noindent
The event samples finally consist of $15\,388$ jets reconstructed in $11\,801$
events for ``high'' \etj 
\linebreak and $26\,848$ jets reconstructed in $21\,001$ events
for ``low'' \etj.
For both samples, the total background of around $2\,\%$ was subtracted. 
The inefficiency due to selection cuts is below $3\,\%$.

\subsection{Cross section determination}
\label{crossdet}

To obtain the inclusive jet cross section, each of the $N$ jets reconstructed
in a given range 
is assigned a weight calculated as the
inverse of the event-by-event trigger efficiency $\epsilon$. The cross section
obtained at the detector level is then corrected by a factor $\mathcal{C}$ for
inefficiencies due to the selection procedure 
and for migrations caused by the detector response.
For instance, the double-differential cross section averaged over 
a range $\Delta \etj  \Delta \etaj$, is defined as:
\begin{equation}
\label{sigmadiff}
\frac{{\rm d}^2 \sigma^{ep\rightarrow e + {\rm jet} + X}}{{\rm d}\etj{\rm d}\etaj} = \frac{\sum_{i=1}^{N}(\frac{1}{\epsilon_{i} })}{\Delta \etj \, \Delta \etaj \, \cal{C} \, \cal{L}}
\end{equation}
where $\cal{L}$ is the integrated luminosity.

\noindent
The trigger efficiency $\epsilon$ was determined from data by using events triggered independently.
For the ``high'' \etj ~analysis, 
$\epsilon$ was parameterised as a function of the 
\ET\ and $\eta$ of the leading jet and 
was always greater than $80\,\%$, reaching $100\,\%$ for $E_T \geq 35~\gev$.
For the ``low'' \etj ~analysis, $\epsilon$
was found to depend on the multiplicity of CT tracks in the event, with
a mean value of $97\,\%$.

\noindent
Two Monte Carlo programs (section~\ref{phenoMC}) were used to correct the data
from each event sample for detector effects. 
For the ``high'' \etj ~sample, HERWIG and PYTHIA were chosen. 
A reasonable description of the observed energy flow around the jet axis
was obtained with both programs, 
provided the underlying event or multiple interaction mechanisms were included in the Monte Carlo simulations~\cite{Ferron:2001}. 
For the ``low'' \etj ~sample, PHOJET, which has been shown to give the best description of energy flow distributions~\cite{Aid:1995ma} and jet profiles~\cite{Ferron:2001}, was chosen
together with PYTHIA.
The mean correction factors $\mathcal{C}$ calculated for each measurement interval
with the two Monte Carlo models were found to lie between $0.9$ and $1.6$
for the ``high'' \etj ~analysis and between $0.5$ and $1.6$ for the ``low'' \etj
~analysis.

\noindent
The bin-to-bin migrations are important due to the steeply-falling shape of the
\etj ~spectrum.
The bin widths were chosen to measure cross sections in as many intervals as possible whilst ensuring that stability and purity\footnote{
The stability $\mathcal{S}$ (purity $\mathcal{P}$) is defined as the number of
jets which are both generated and reconstructed in an analysis bin, divided by
the total number of jets that are generated (reconstructed) in that bin. By
definition, $\mathcal{C} \equiv \mathcal{S} / \mathcal{P}$.}
were greater than $30 \,\%$.
These criteria restrict the \etaj ~cross section measurements to different \etaj ~ranges depending on the \etj ~range considered.
At ``high'' \etj, the problematic region is
that of negative \etaj.
Due to the Lorentz boost between the $\gamma p$ cms and the laboratory frame, $\etaj = 0$ corresponds to a cms pseudorapidity\footnote{
$\eta ^\star = \eta - \ln(2 E_p/ W_{\gamma p})$.}
 $\eta ^\star$ of about $-2$, which is well into the photon hemisphere.
The cross section thus falls most steeply with increasing \etj ~in this region.
At ``low'' \etj, the influence of the proton remnant compromises measurements towards higher \etaj.

\subsection{Systematic uncertainties}

The following systematic uncertainties have been considered :
\begin{itemize}
\item The uncertainty in the absolute hadronic energy scale of the LAr calorimeter ($2\,\%$ for ``high'' \etj ~and $4\,\%$ for ``low'' \etj) leads to an uncertainty of typically $10\,\%$ for ``high'' \etj ~and between $10\,\%$ and $20\,\%$ for ``low'' \etj.
\item The $7\,\%$ uncertainty in the hadronic SPACAL energy scale results in an uncertainty of $1\,\%$ at ``high'' \etj ~and is negligible at ``low'' \etj.
\item The $3\,\%$ uncertainty in the fraction of the energy of the reconstructed hadronic final state carried by tracks leads to an uncertainty of less than $1\,\%$ for ``high'' \etj ~and of $2\,\%$ to $4\,\%$ for ``low'' \etj.
\item The background subtraction leads to an uncertainty of $1\,\%$.
\item The statistical uncertainty in the trigger efficiency determination leads to an uncertainty
of $1 \,\%$ or less.
\item The uncertainty in the integrated luminosity results in an overall normalisation error of $1.5\,\%$.
\item The uncertainty in the correction for detector effects was taken to be half the difference between the correction factors calculated from the two Monte
Carlo programs. 
It is smaller than $8\,\%$ for ``high'' \etj ~and smaller than $10\,\%$ for ``low'' \etj.
\end{itemize}

\noindent
All systematic uncertainties are added in quadrature. The resulting uncertainty
ranges from $10\,\%$ to $20\,\%$ for  ``high'' \etj ~and from $15\,\%$ to $30\,\%$ for 
``low'' \etj ~and is dominated by the normalisation uncertainty due to the LAr calorimeter energy 
scale.

\section{Results}
\label{result}

\begin{figure}[h]
\begin{center}
\includegraphics[width=0.84\textwidth]{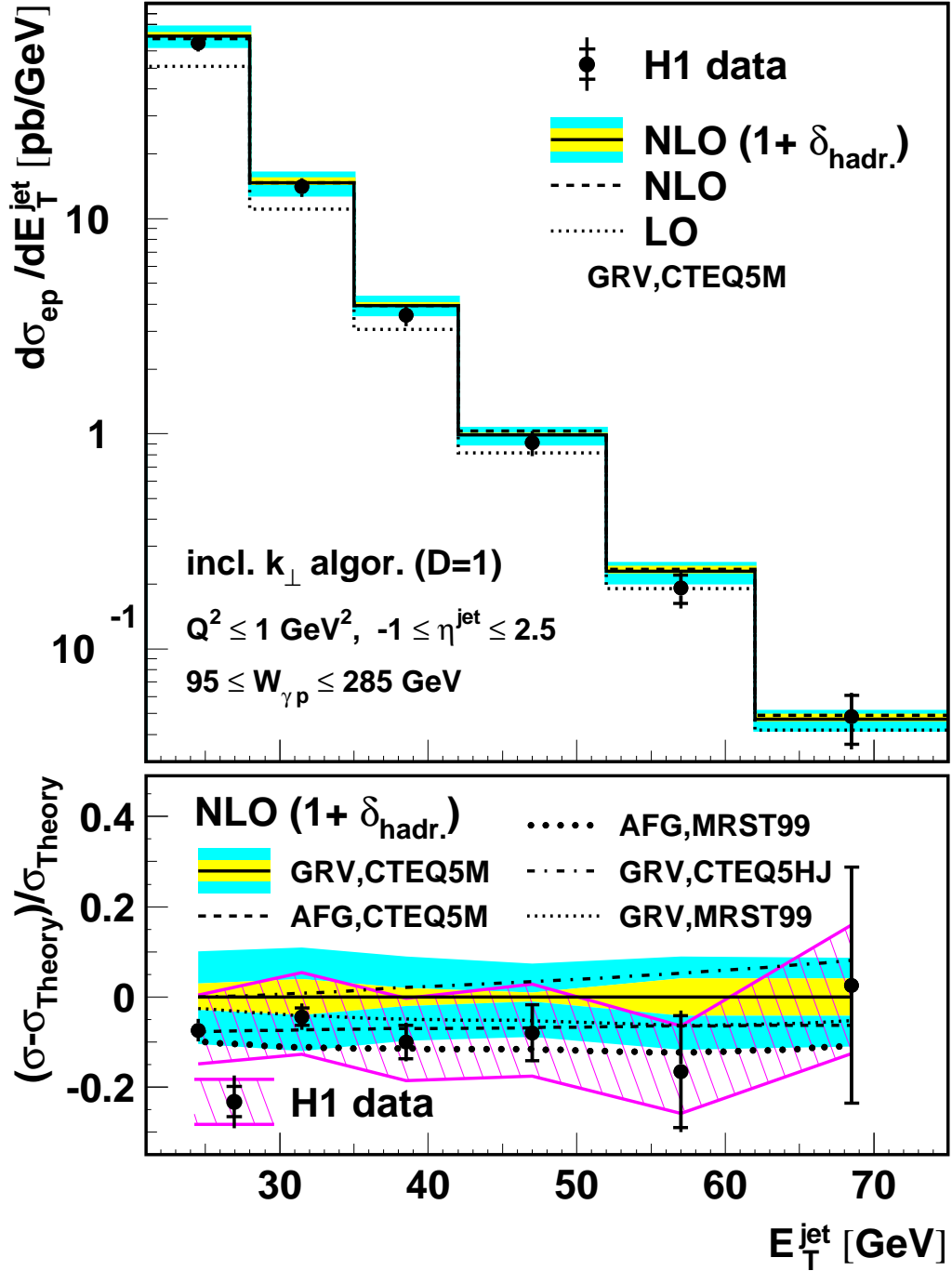}
\end{center}
\it \caption{\label{fig:fullinclhetet}Top: differential $e^+ p$ cross section for inclusive jet production as a function of \etj ~integrated over $-1 \leq \etaj \leq 2.5$.
The data are compared with LO and NLO QCD calculations using GRV photon PDFs and CTEQ5M proton PDFs.
Bottom: relative difference between the data or different calculations and the NLO prediction with hadronisation corrections.
The uncertainty associated with the LAr energy scale is shown as a hatched band.
The shaded band displays the uncertainty on the NLO QCD predictions.
The inside part shows the uncertainty associated with the hadronisation corrections, the outside part shows the uncertainty associated with the choice of the renormalisation and factorisation scales and both uncertainties are added linearly. 
}
\end{figure}

\begin{figure}[b]
\begin{center}
\includegraphics[width=\textwidth]{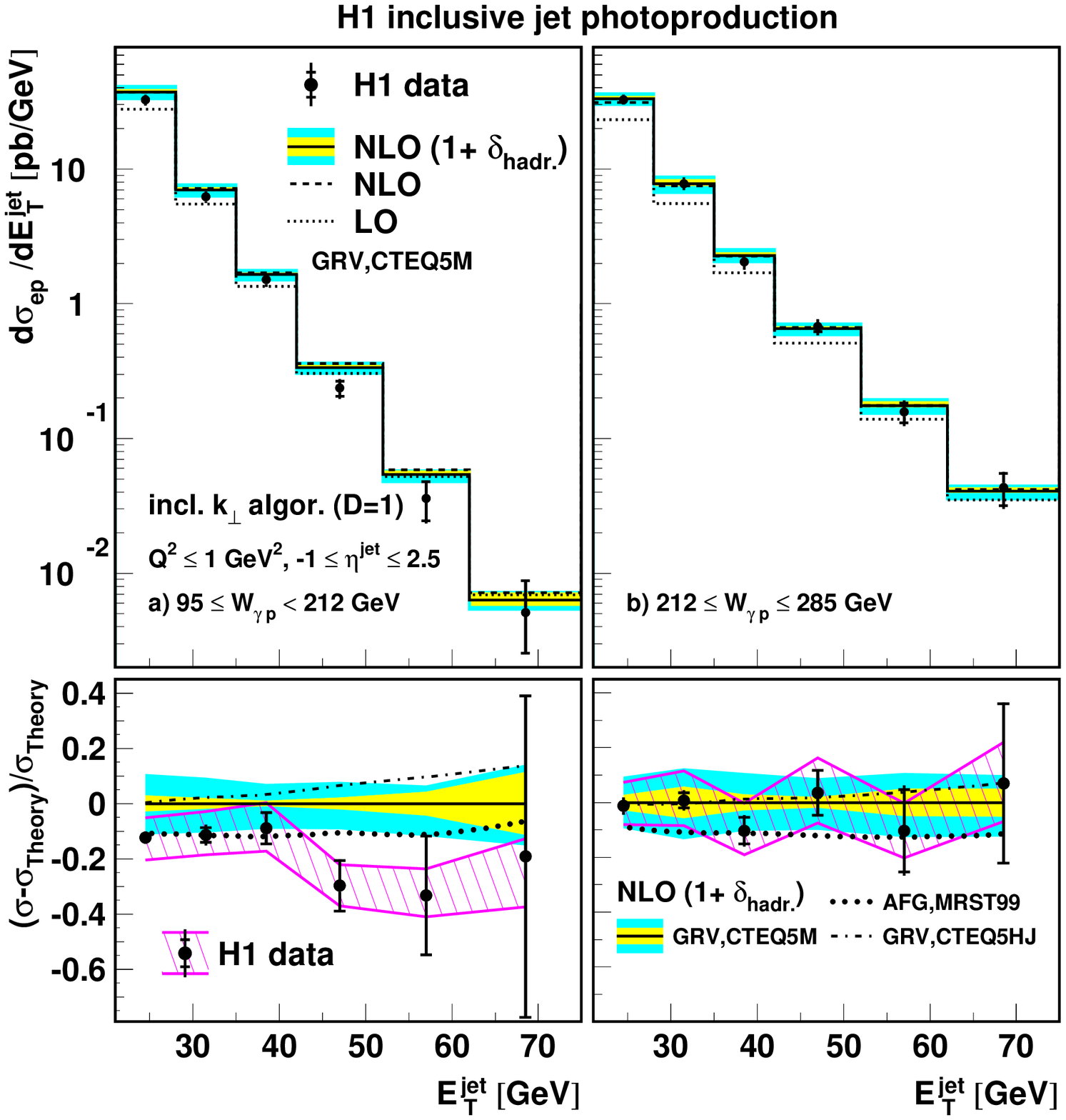}
\end{center}
\it \caption{\label{fig:yevolinclhetet}Top: differential $e^+ p$ cross section for
inclusive jet production as a function of \etj ~integrated over $-1 \leq \etaj
\leq 2.5$ for  $95 \leq \Wgp <  212~\gev$ (a) and $212 \leq \Wgp \leq 285~\gev$  (b). 
Bottom: relative difference between the
data or different calculations and the NLO calculation, including  hadronisation corrections, based on GRV and CTEQ5M 
(see Fig.~\ref{fig:fullinclhetet} caption for further details). 
}
\end{figure}

\begin{figure}[h]
\begin{center}
\includegraphics[width=0.87\textwidth]{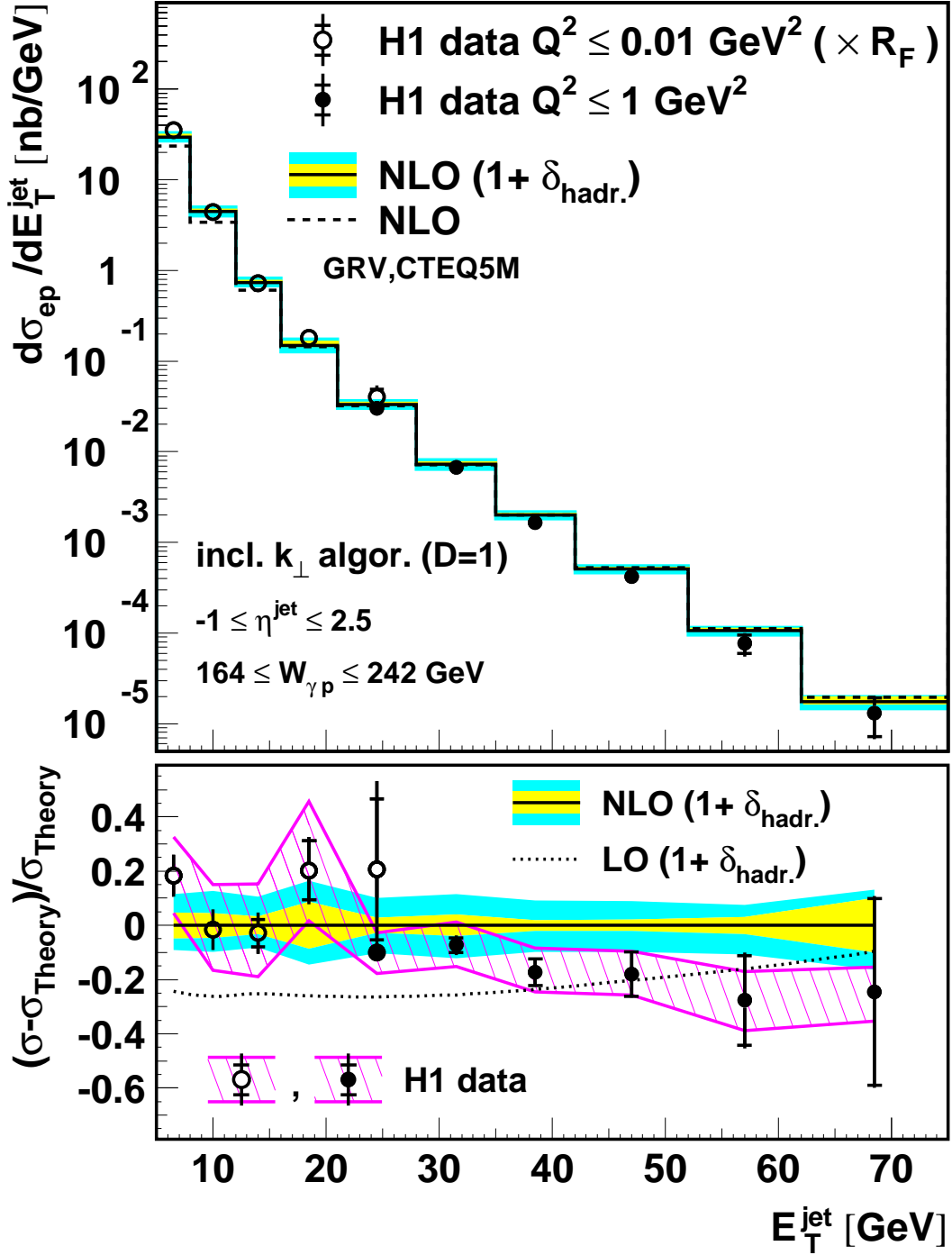}
\end{center}
\it \caption{\label{fig:allETinclkT}Top: differential $e^+ p$ cross section for
inclusive jet production as a function of \etj ~integrated over $-1 \leq \etaj
\leq 2.5$ and $Q^2 \leq 1~\gevsq$. The ``low'' \etj ~part, measured 
for $Q^2 \leq 0.01~\gevsq$, is corrected by a factor $R_F$ which is 
the ratio of the photon fluxes in the two $Q^2$ regions (see text). 
The photon-proton centre-of-mass energy is restricted to the range
$164 \leq \Wgp \leq 242~\gev$. The data stemming from the analysis at
``low'' and ``high'' \etj ~are indicated by empty and full points respectively. 
Bottom: relative difference between the
data or LO QCD prediction and the NLO calculation, including hadronisation corrections, based on GRV and CTEQ5M (see Fig.~\ref{fig:fullinclhetet} caption for further details).
}
\end{figure}

\begin{figure}[h]
\begin{center}
\includegraphics[width=0.98\textwidth]{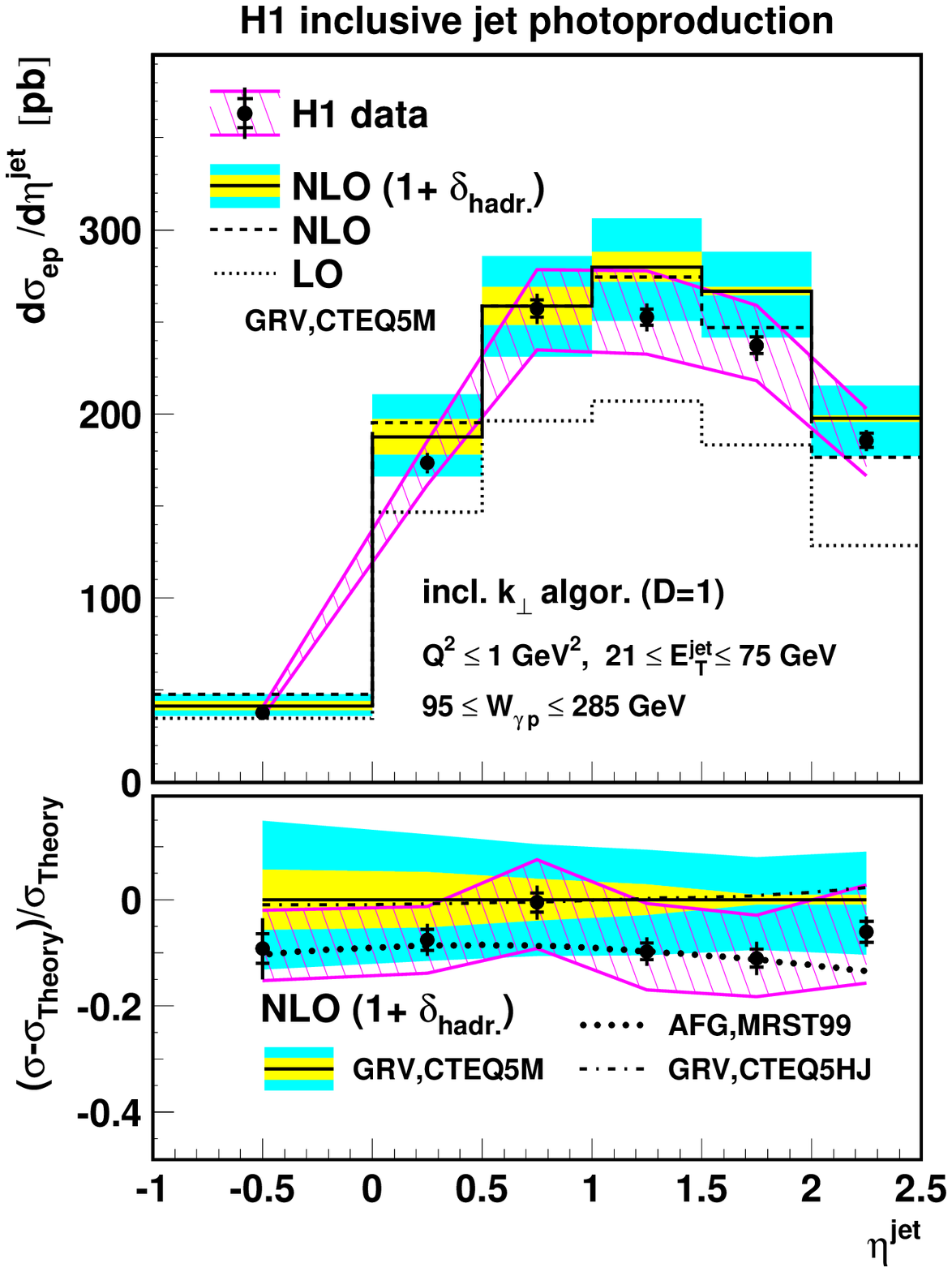}
\end{center}
\it \caption{\label{fig:fullinclhetrp}Top: differential $e^+ p$ cross section for inclusive jet production as a function of \etaj ~integrated over $21 \leq \etj
\leq 75~\gev$. 
Bottom: relative difference between the data or different calculations and the NLO calculation, including  hadronisation corrections, based on GRV and CTEQ5M 
(see Fig.~\ref{fig:fullinclhetet} caption for further details). 
} 
\end{figure}

In this section, inclusive jet cross sections are presented,
corrected for detector effects and measured in different kinematic regions as functions of \etj ~and \etaj ~in the laboratory frame.
Good agreement with previous
data~\cite{Abt:1993mx,Derrick:1994aq,Aid:1995ma,Breitweg:1998ur}
has been found when using the cone algorithm~\cite{Ferron:2001}. The results 
are also consistent with recently published ZEUS data~\cite{unknown:2002ru} using the $k_{\bot}$ algorithm.
The numerical results using the $k_{\bot}$ algorithm 
(Tabs.~\ref{tab:fullinclhetet} to~\ref{tab:fullinclhetrplow})
are given 
as differential cross sections averaged over the quoted ranges.
Those obtained with the cone algorithm (Tabs.~\ref{tab:combfullinclhetetco} and~\ref{tab:combxt}) are given at the average value in each analysis interval, determined according to the Monte Carlo simulation.
The results are shown in Figs.~\ref{fig:fullinclhetet} 
to~\ref{fig:fxT}.
In the \etj ~spectra (upper part of Figs.~\ref{fig:fullinclhetet} to~\ref{fig:allETinclkT}), all systematic uncertainties are added in quadrature with the statistical uncertainty and are shown as error bars.
The inner error bars denote the statistical and the outer the total uncertainty.
In all other figures, the LAr calorimeter energy scale uncertainty is not included in the error bars, but is shown separately as a hatched band. 
The results are compared with LO and NLO QCD predictions first, then with similar measurements at $p\overline{p}$ colliders.
Unless otherwise stated, the QCD prediction is calculated at NLO with the standard settings described in section~\ref{phenoNLO} and corrected for hadronisation, as explained in section~\ref{hadronisation}.
Its uncertainty 
is shown as a shaded band divided into two parts.
The inside (light) part is the uncertainty associated with the hadronisation
corrections and the outside (dark) part is the uncertainty associated with the
choice of the renormalisation and factorisation scales.
These uncertainties are added linearly.
When presented (Figs.~\ref{fig:fullinclhetet} to~\ref{fig:fullinclhetrp}), 
the relative differences are always defined with respect to this standard NLO QCD prediction.
The total hadronisation correction factors $(1+\delta_{hadr.})$ and their errors are given in the tables, together with the correction factors associated with the effects of fragmentation, $(1+\delta_{frag.})$ and of the underlying event, $(1+\delta_{u.e.})$.

\subsection{{\boldmath \etj} ~cross sections}
\label{result:etj}
The measured differential $e^+ p$ cross section \setj ~for inclusive jet production integrated over $-1 \leq \eta^{jet} \leq 2.5$ in the kinematic region defined by $Q^2 \leq 1~\gevsq$ and $95 \leq \Wgp \leq 285~\gev$ is shown for $\etj > 21~\gev$ in Fig.~\ref{fig:fullinclhetet}~(top).
The LO QCD calculation fails to reproduce the normalisation of the distribution.
Both NLO predictions, with and without hadronisation corrections, reproduce the measured distribution very well.
As shown in Fig.~\ref{fig:fullinclhetet}~(bottom), the uncertainty due to the renormalisation and factorisation scales is of the order of $10\,\%$.
The calculated cross sections using the GRV photon PDFs are typically $5\,\%$~to~$10\,\%$ larger than those obtained with AFG.
To show the sensitivity to the proton PDFs, the predictions using GRV for the photon and MRST99 or CTEQ5HJ for the proton are also shown.
Compared with CTEQ5M, MRST99 and CTEQ5HJ give almost the same prediction at relatively small \etj, but show differences as \etj ~increases.
The prediction using MRST99 decreases relative to that using CTEQ5M by $5\,\%$ over the measured \etj ~range while that using CTEQ5HJ increases by $8\,\%$.
Within the errors,
the NLO QCD calculations with each of the PDFs choices describe the magnitude and the shape of the measured inclusive \etj ~spectrum very well, up to the highest measured \etj ~values.
 
\noindent
In Fig.~\ref{fig:yevolinclhetet}~(top), \setj ~is presented for two separate \Wgp ~intervals
$95 \leq \Wgp <  212~\gev$ and $212 \leq \Wgp \leq 285~\gev$.
%The corresponding \Wgp ~mean values 
%are $\mWgp \simeq 130~\gev$ 
%and  $\mWgp \simeq 250~\gev$.
At higher \Wgp, the \etj ~spectrum is harder and extends to higher \etj ~values, as expected.
The predictions of the two combinations of photon and proton PDFs which give the lowest (AFG for the photon and MRST99 for the proton) and the highest (GRV for the photon and CTEQ5HJ for the proton) cross sections are also shown.
As can be seen in Fig.~\ref{fig:yevolinclhetet}~(bottom),
all these NLO QCD calculations describe 
the magnitude and the shape of the inclusive \etj ~spectra measured in the two 
\Wgp ~ranges.

\noindent
A measurement of the inclusive jet cross section over the whole \etj ~range was
performed by combining the ``low'' and ``high'' \etj ~data samples.
In order to do this, the same \Wgp ~cut was applied to the ``high'' as to the ``low'' \etj ~data sample, i.e. $164\leq \Wgp \leq 242~\gev$. 
%resulting in a mean $\gamma p$ cms energy $\mWgp = 200~\gev$.
The ``low'' \etj ~cross section was also corrected 
to correspond to the same $Q^2$ range\footnote{
The ``low'' \etj ~cross section was multiplied by the ratio $R_F = F(1~\gev ^2) / F(0.01~\gev ^2)$, where $F(\qs_{max})$ is the integral of the photon flux $f_{\gamma /e}(y,\qs)$ (see Eqs.~\ref{sigmaep_dir} and~\ref{sigmaep_res}) 
over $\qs < \qs_{max}$  
in the range $0.3 \leq y \leq 0.65$, which corresponds to the chosen \Wgp ~range.
The numerical integration yields $F(1~\gev ^2) = 0.0181$, $F(0.01~\gev ^2) = 0.0127$ and $R_F = 1.43$.
}
\setcounter{fndeux}{\thefootnote}
as the ``high'' \etj ~sample.
The measured cross section is shown in Fig.~\ref{fig:allETinclkT}.
The analyses of the ``low'' and ``high'' \etj ~samples agree well in their domain of overlap.
The measured \setj ~cross section falls by more than 6 orders of magnitude between $\etj=5$ and $75~\gev$ and is well reproduced by the theoretical prediction. 
The NLO contribution and the hadronisation corrections are both needed to give a good agreement of the calculation with the measured data.

\noindent
Following a procedure previously applied to the inclusive charged particle photoproduction cross section~\cite{Adloff:1998vt}, the power-law~\cite{hagedorn} function $\propto (1+\etj/E_{T,0})^{-n}$ was fitted to the inclusive jet cross section $1 / \etj \cdot {\rm d} \sigma / {\rm d} \etj$ obtained from the data of Fig.~\ref{fig:allETinclkT}.
Good fits with
stable results could only be obtained in the region $ 5 \leq \etj < 35 ~
\gev$. 
The fit 
gave the results
$E_{T,0} = 2.4\pm 0.6 \, (\rm{stat.}) ~^{+ 0.2}_{- 0.6}\, (\rm{syst.}) ~\gev$ and $ n = 7.5\pm 0.3 \, (\rm{stat.}) ~^{+ 0.1}_{- 0.5} \, (\rm{syst.})$.
The value of the exponent is in agreement with that obtained in~\cite{Adloff:1998vt}: $n = 7.03 \pm 0.07 \, (\rm{stat.}) ~\pm 0.20 \, (\rm{syst.})$.

\subsection{{\boldmath \etaj} ~cross sections}
The measured differential $e^+ p$ cross section \setaj ~in the range $-1 \leq \etaj \leq 2.5$ for inclusive jet production integrated over $21 \leq \etj \leq 75~\gev$, $Q^2 \leq 1~\gevsq$ and $95 \leq \Wgp \leq 285~\gev$ is shown in Fig.~\ref{fig:fullinclhetrp}.
The calculated cross sections using different combinations of the photon and proton PDFs give a good description of the data, within the experimental and theoretical errors.
The normalisation of the data can only be described when the NLO contribution is included.
The description is not significantly improved by the hadronisation corrections.

\begin{figure}[h]
\begin{center}
\includegraphics[width=\textwidth]{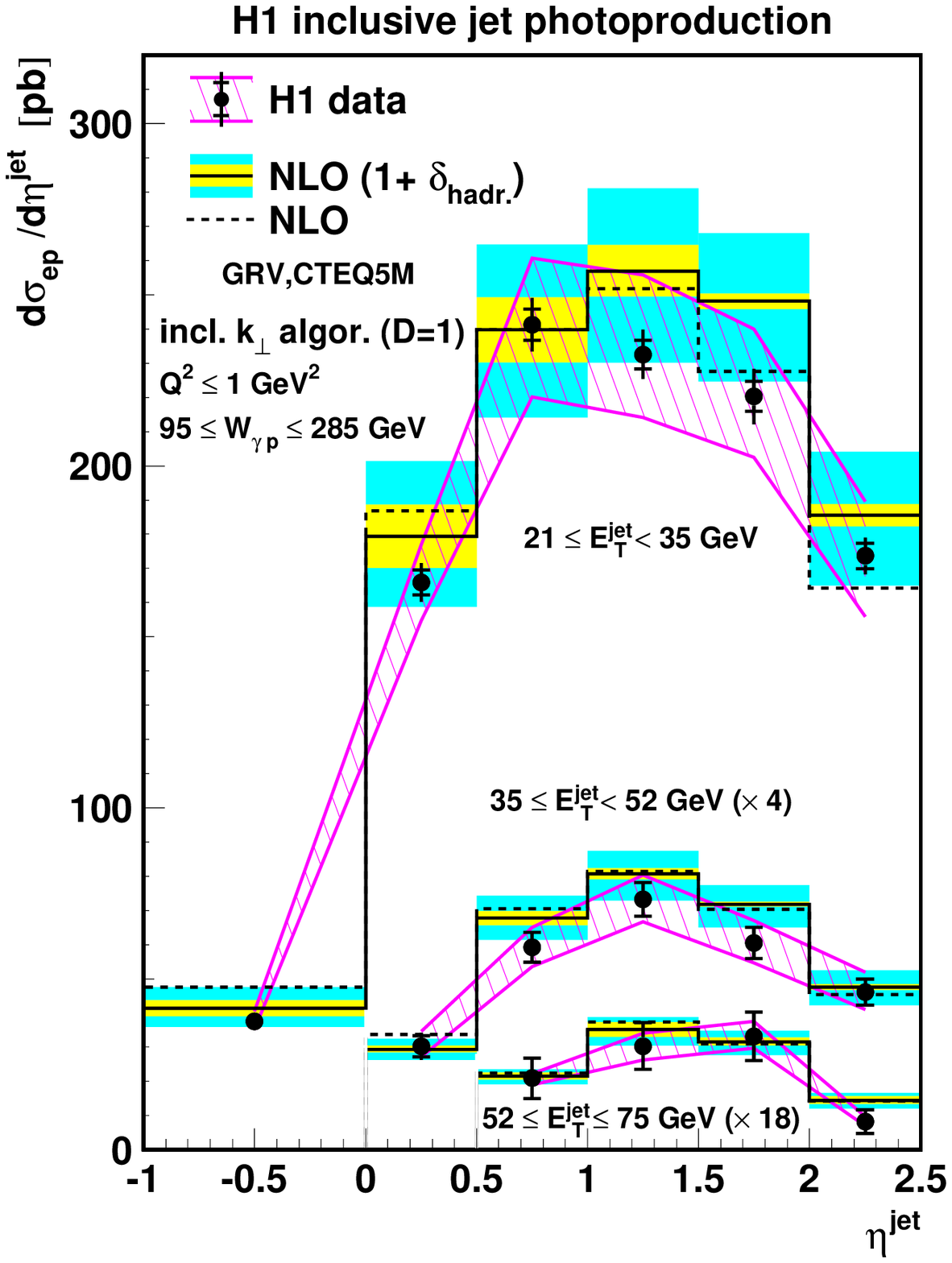}
\end{center}
\it \caption{\label{fig:etinclhietrp}Differential $e^+ p$ cross section for
inclusive jet production as a function of \etaj ~integrated over various \etj
~ranges.  
The data are compared with NLO QCD predictions obtained  by
using GRV photon PDFs and CTEQ5M proton PDFs (see Fig.~\ref{fig:fullinclhetet}
caption for further details). 
} 
\end{figure}

\noindent
In Fig.~\ref{fig:etinclhietrp},
\setaj ~is presented in three different intervals of \etj
~and compared with NLO QCD predictions.
The hadronisation corrections correspond to an increase (decrease) of the
pure partonic prediction in the forward (backward) region.
The increase in the forward region is due to the influence of the proton remnant leading to significant underlying energy, while 
the decrease in the backward region
reflects the escape of partonic energy from the jet due to fragmentation (section~\ref{hadronisation}).
Within the errors, the data are well described by the NLO QCD predictions.

\begin{figure}[h]
\begin{center}
\includegraphics[width=\textwidth]{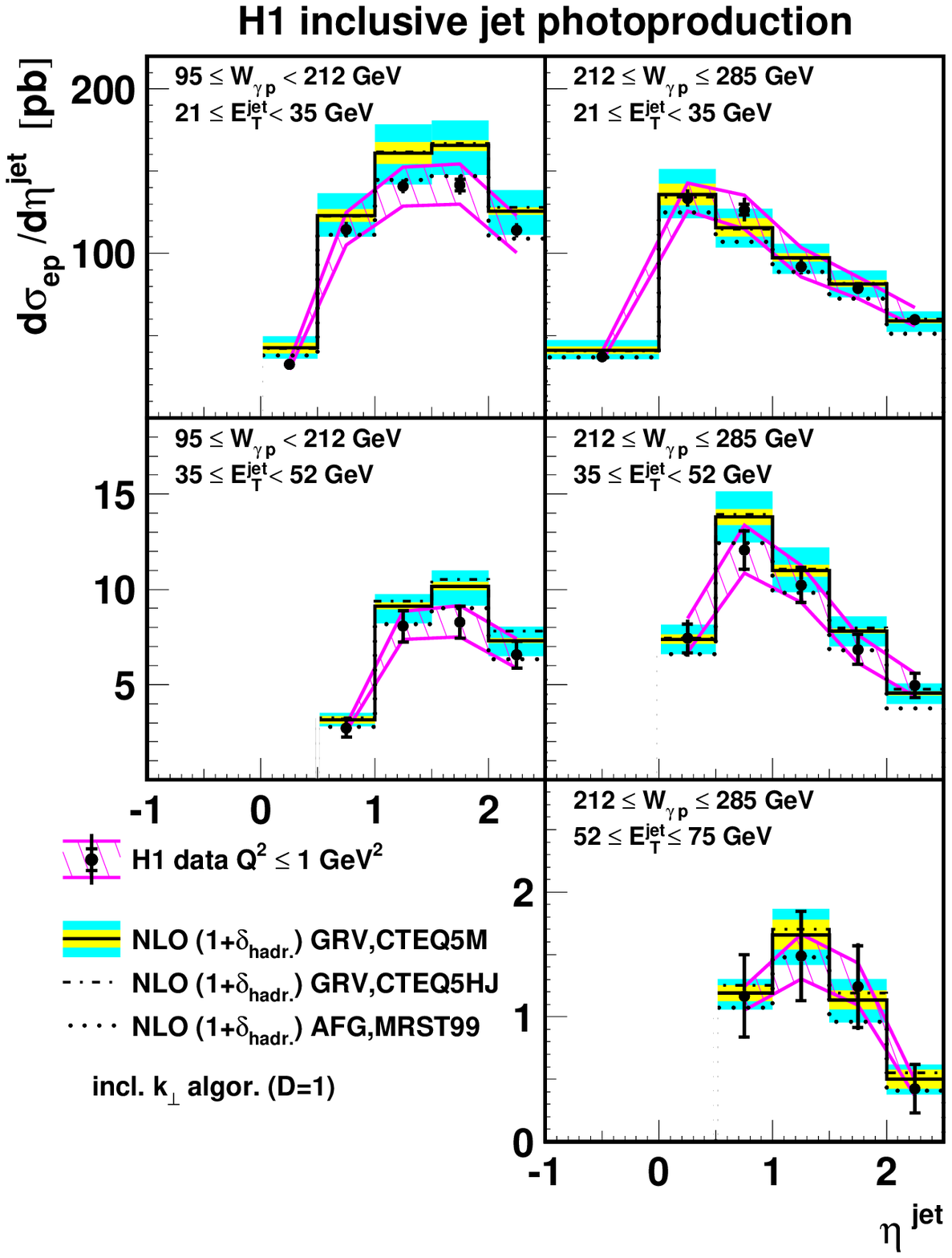}
\end{center}
\it \caption{\label{fig:inclhetrp}Differential $e^+ p$ cross section for
inclusive jet production  as a function of \etaj ~integrated over various \etj
~and \Wgp ~ranges. 
The data are compared with NLO QCD predictions obtained by using different photon
and proton PDFs (see Fig.~\ref{fig:fullinclhetet} caption for further details). 
} 
\end{figure}

\noindent
In order to study the cross section more differentially, measurements of \setaj ~in two regions of \Wgp ~and three intervals of \etj ~are presented in Fig.~\ref{fig:inclhetrp}.
The maximum of the cross section is shifted towards low \etaj ~values at higher \Wgp ~due to the 
decreased forward boost of the hadronic cms relative to the laboratory frame.
The NLO QCD predictions with and without (not shown in Fig.~\ref{fig:inclhetrp}) hadronisation corrections are in general in good agreement with the measured cross sections. 
NLO QCD calculations using different combinations of photon and proton PDFs also give good agreement with the data.
The precision of the measurement as well as the theoretical uncertainties 
do not allow any firm conclusion to be drawn on which combination of PDFs is favoured by the data.

\begin{figure}[h]
\begin{center}
\includegraphics[width=\textwidth]{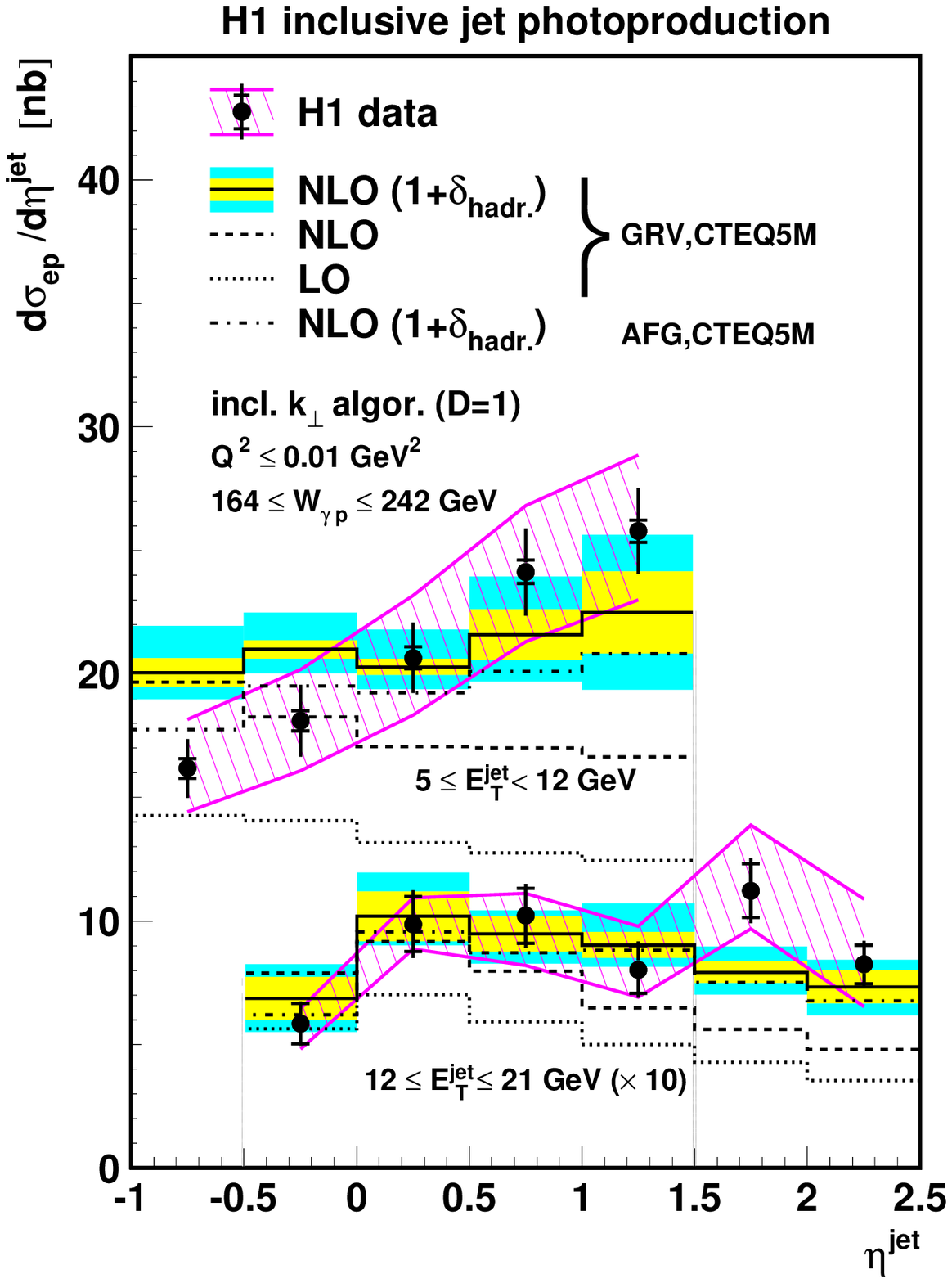} 
\end{center}
\it \caption{\label{fig:etinclloetrp}Differential $e^+ p$ cross section for
inclusive jet production as a function of \etaj ~integrated over various \etj
~ranges. 
The data are compared with LO and NLO QCD predictions obtained  by
using GRV or AFG photon PDFs and CTEQ5M proton PDFs (see Fig.~\ref{fig:fullinclhetet}
caption for further details). 
} 
\end{figure}

\noindent
The \setaj ~measurement for the ``low'' \etj ~sample is presented in Fig.~\ref{fig:etinclloetrp} for two intervals of \etj, in the
kinematic region $\qs \leq {10}^{-2}~\gevsq$ and $164 \leq \Wgp \leq 242~\gev$.
The NLO predictions are in agreement with the data in the range $12 \leq \etj \leq 21 ~\gev$, provided the hadronisation corrections are included.
In the lowest \etj ~range $5 \leq \etj < 12 ~\gev$, however, the  agreement with the NLO predictions including hadronisation corrections is marginal and the data seem to indicate  a rise of the cross section with increasing \etaj ~which is faster than in the theoretical predictions. 
This may be the result of a failure of the LO Monte Carlo to accurately estimate the hadronisation corrections. Inadequacy of the photon PDFs in this kinematic
range or the absence of higher order corrections beyond NLO may also be responsible.

\subsection{Comparison with 
{\boldmath $p\overline{p}$} collider results}
\label{result:ppbar}
It is interesting to compare the present photoproduction measurements with similar $p \overline{p}$ data 
in order to see the effects arising from the different structure of the photon and the proton.
The differential $e^+p$ cross section \setj
~was measured in the range $-1 \leq \etaj \leq 2.5$, 
as in Fig.~\ref{fig:allETinclkT}, but using the cone algorithm with cone radius $R=1$ to match the procedure used for the only available $p \overline{p}$ 
data~\cite{Albajar:1988tt} at comparable cms 
energy $\sqrt{s} = 200 ~\gev$. 
The results 
are presented in Tab.~\ref{tab:combfullinclhetetco}.
%, together with the 
%average \etj ~value for each interval, as given by PYTHIA.
Up to a normalisation factor, the
\etj ~dependence of the data is compatible in the region of 
overlap $5 \leq \etj \leq 22 ~\gev$ with that of~\cite{Albajar:1988tt}.
Monte Carlo studies show that the difference in
cms jet pseudorapidity ranges between the
$\gamma p$ ($-3.0 \stackrel{<}{_{\sim}} 
\eta^\star \stackrel{<}{_{\sim}} 0.5$) and 
$p \overline{p}$ ($| \eta^\star | < 1.5$) 
data does not affect this conclusion.
%According to PYTHIA, changing this $\eta^\star$ range to that of the $e^+ p$ data would 
%lead to a decrease of the $p \overline{p}$ cross section of between $25\,\%$ and $35\,\%$
%and so wouldn't change this conclusion.

\noindent
To allow comparisons with $p \overline{p}$ 
measurements at different energies, the scaled $\gamma p$ cross section 
\begin{equation}
\label{sigmafxt}
S(x_T)  \ \equiv \ E_T^{\,jet ^{\, \scriptstyle 4}} \, E^{\,jet} \, \frac{ {\rm d}^3 \sigma}{ {\rm d}p^{\, jet ^{\, \scriptstyle 3}}} \ = \ \frac{E_T^{\,jet ^{\, \scriptstyle 3}}}{2 \pi} \, \frac{ {\rm d}^2\sigma} { {\rm d}\etj {\rm d}\etaj } \ ,
\end{equation}
where $(E^{\,jet},p^{\, jet})$ is the four-vector of the jet,
was measured as a function of the dimensionless variable $x_T = 2 \etj / \Wgp$.
In the naive parton model, $S(x_T)$ is independent of cms energy for the same 
colliding particles.
%To extract $S(x_T)$, t
The differential $e^+ p$ cross section \setj ~was measured 
with the cone algorithm in the restricted range $1.5 \leq \etaj \leq 2.5$ 
and $\etj > 8 ~\gev$. 
It was then transformed into 
$S(x_T)$ at a fixed $\Wgp = 200~\gev$ averaged over the range 
$| \eta ^\star | \leq 0.5$ using the Monte Carlo models to evaluate the 
correction factors and their uncertainties.

\begin{figure}[b] \unitlength 1mm
\begin{center}
 \begin{picture}(80,180)
   \put(-43,5){\epsfig{file=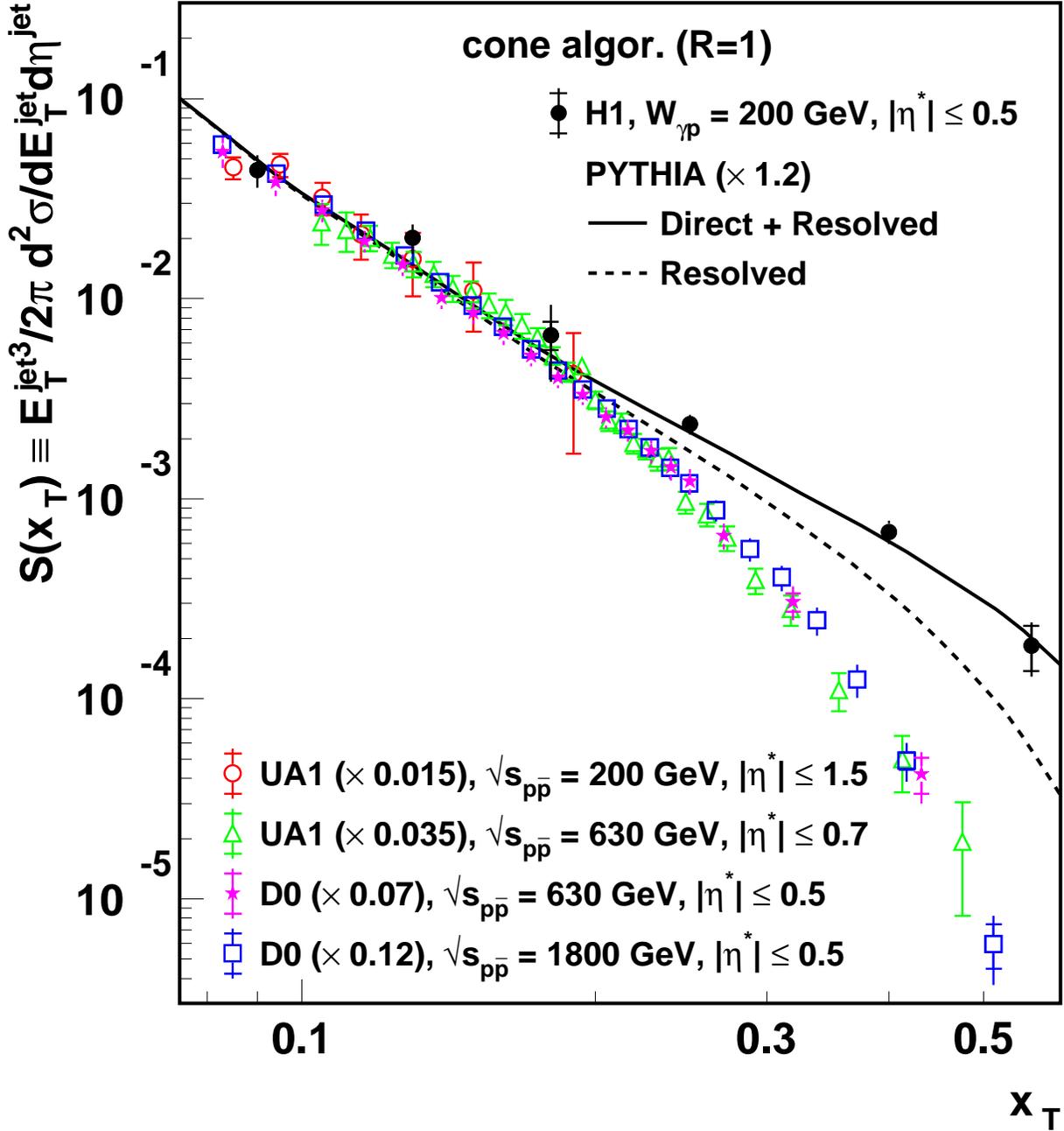,width=\textwidth}}
%   \put(-1,15){\bf {\Large 0.1}}
%   \put(69,15){\bf {\Large 0.3}}
 \end{picture}
\end{center}
\vspace*{-0.3cm}
\it \caption{\label{fig:fxT}Scaled $\gamma p$ cross section at 
$\Wgp = 200 ~\gev$ 
for inclusive jet production as a function of $x_T$ for 
$|\eta^\star| \leq 0.5$. Jets are found with the cone algorithm ($R=1$).
The data are compared with measurements
from UA1~\cite{Arnison:1986vk,Albajar:1988tt} and D0~\cite{Abbott:2001ce,Abbott:1998ya} 
 of inclusive jet production  in $p\overline{p}$
collisions at various cms energies.
The predictions of PYTHIA for $\gamma p$ and for the resolved photon contribution are also shown, multiplied by a factor $1.2$ to match the normalisation
of the data.
}
\end{figure}

\noindent
The $S(x_T)$ distribution is presented in Tab.~\ref{tab:combxt} and
Fig.~\ref{fig:fxT}. In the figure, it is compared
with data from $p \overline{p}$ scattering obtained by the 
UA1~\cite{Arnison:1986vk,Albajar:1988tt} and 
D0~\cite{Abbott:2001ce,Abbott:1998ya} collaborations at various energies 
using the cone algorithm. The $p \overline{p}$ data were transformed into 
$S(x_T)$ using the \etj ~value at 
the centre of each measurement interval and were scaled by factors of 
${\cal O}(\alpha_{em}/\alpha_S)$ such that $S(x_T)$ approximately matches
that from the photoproduction data at $x_T \sim 0.1$. 
%The UA1~\cite{Arnison:1986vk,Albajar:1988tt} and D0~\cite{Abbott:2001ce,Abbott:1998ya} measurements at
%various energies were transformed
%into $S(x_T)$ by using \etj ~at the centre of each measurement interval and were normalised
%using factors 
%${\cal O}(\alpha_{em}/\alpha_S)$.
%The scaled cross section $S(x_T)$
%is presented in Fig.~\ref{fig:fxT}. 
Despite the differences in the $\eta^\star$ ranges of measurement and in the details of the analysis procedure\footnote{
UA1 measured cross sections in the range $|\eta^\star| \leq 1.5$ for $\sqrt{s}=200~\gev$ and $|\eta^\star| \leq 0.7$ for $\sqrt{s}=630~\gev$, using a cone radius $R = 1$ and no jet pedestal energy subtraction. D0 measured cross sections in the range $|\eta^\star| \leq 0.5$ using a cone radius $R = 0.7$ and jet pedestal energy subtraction.}, 
%all $p \overline{p}$ data are compatible in shape to better than $50 \, \%$.
all $p \overline{p}$ data are in approximate agreement after the
scaling factors are applied.
Within the experimental uncertainties, the shape of the 
$S(x_T)$ distribution for $\gamma p$ is 
compatible with those from $p \overline{p}$ data 
in the region $x_T \lesssim 0.2$,
where the resolved photon leads to a similar behaviour  
of the scaled cross section to that for a 
hadron except for the overall normalisation. 
At larger
$x_T$, the shape of the $\gamma p$ cross section begins to deviate 
from that for $p\overline{p}$. 
As can be inferred from the PYTHIA predictions for the 
full $\gamma p$ cross section and for the contribution from resolved photons,
this is due to the 
enhancement of the resolved photon quark density relative to that 
of the proton at large momentum fractions, as well as the increasing proportion of
direct photon interactions. The direct photon contribution involves the convolution of only
one set of PDFs and dominates the scaled cross section at the largest $x_T$.
%This may be interpreted as a confirmation of the dual nature of the photon. 
%For low $x_T$, the photon shows a similar behaviour to that of a hadron, except for the overall normalisation of the cross-section.
%At higher $x_T$, the presence of the
%anomalous\footnote{The anomalous part of the photon structure
%is the perturbative component arising from the coupling $\gamma \rightarrow q \overline{q}$.} and the direct
%components of the photon 
%becomes dominant, as can be seen from the PYTHIA predictions. 
%Because of the anomalous part, the structure of the resolved photon is different from the one of the proton, which enhances the quark density in the photon compared with the proton at large $x$ and large scales and hence the resolved $\gamma p$ cross section compared with the $p\overline{p}$ cross section.
%In contrast to $p\overline{p}$ collisions, the direct
%$\gamma p$ cross section involves the convolution of only one proton PDF
%and then dominates the scaled cross section at the highest $x_T$. The sum of the
%anomalous and direct contributions
%thus gives rise to a harder spectrum in $\gamma p$ 
%than in $p \overline{p}$ collisions  at $x_T$ above 0.2.

\section{Summary}

A new measurement of inclusive jet production cross sections in quasi-real photoproduction ($Q^2 \leq 1~\gevsq$) has been presented, based on an integrated luminosity of $24.1~\pbi$ of 
$e^+ p$ data collected by the H1 experiment in the years $1996$ and $1997$.
Compared with the last published H1 result~\cite{Aid:1995ma} on this topic, this measurement represents an increase in luminosity by a factor of $80$.
The jets were selected using the inclusive $k_{\bot}$ algorithm in the pseudorapidity range $-1 \leq \etaj \leq 2.5$ in the laboratory frame.
The photon-proton centre-of-mass energy range of the measurement for jets 
with transverse energies $ \etj \geq 21~\gev$ is 
$95 \leq \Wgp \leq 285~\gev$.
The measurement could be extended down to $ \etj \geq 5~\gev$ by using a sample of data with integrated luminosity $0.47~\pbi$, 
collected in a 
data taking period with a dedicated trigger.
There, the kinematic range of measurement was  $Q^2 \leq 0.01~\gevsq$ and 
$164 \leq \Wgp \leq 242~\gev$.

\noindent
The measured cross sections were corrected to the hadron level 
and compared with leading order (LO) and next-to-leading order (NLO) QCD calculations, with and without fragmentation and underlying event corrections.
The LO QCD calculations are unable to reproduce the normalisation of the experimental data.
The NLO QCD calculations, using various available photon and proton PDFs, describe the measured distributions both in normalisation and shape over the whole \etj ~and \etaj ~range within the experimental and theoretical uncertainties.
%This result is consistent with recently published dijet cross section measurements~\cite{Adloff:2002au}.
For $\etj \geq 21 \, \gev$, the hadronisation corrections to the NLO QCD calculations only slightly improve the agreement with the data, 
whereas for $5 \leq \etj < 21 \, \gev$, good agreement can only be obtained with the hadronisation corrections.
The current precision of the experimental results as well as of the theoretical predictions does not allow one to discriminate between the different photon and proton PDFs using these data alone.
However, the information obtained from these measurements could be used to constrain 
the photon and proton PDFs in global fits of experimental results.

\noindent
To compare with previous measurements at HERA and at $p\overline{p}$ colliders, the inclusive \etj ~differential cross section was also measured for jets defined using the cone algorithm with $R = 1$.
The shape of the $\gamma p$ scaled cross section $S(x_T)$, as a function of the dimensionless variable $x_T = 2 \etj / \Wgp$ at $\Wgp = 200 \ {\rm GeV}$ for
$| \eta^\star | < 0.5$, is compatible with that of similar $p\overline{p}$ measurements for $x_T \lesssim 0.2$. The shapes for $\gamma p$ and $p\overline{p}$ are different at larger
$x_T$, where resolved photon structure at large $x_\gamma$ and direct photon
interactions become important.  
%The difference in colliding particle structure arising from the $\gamma \rightarrow q \overline{q}$ branching provides a natural explanation for the deviation of the $\gamma p$ with respect to the $p\overline{p}$ cross section observed at higher $x_T$. 

\section*{Acknowledgements}
We are grateful to the HERA machine group whose outstanding efforts have made 
this experiment possible.
We thank the engineers and technicians for their work in constructing and now maintaining the H1 detector, our funding agencies for financial support, the DESY technical staff for continual assistance and the DESY directorate for 
support and for the hospitality which they extend to the non DESY  members of the collaboration.
We wish to thank M.~Fontannaz for valuable discussions and S.~Frixione and
G.~Ridolfi for making their theoretical calculations available to us.

\clearpage

\renewcommand{\tabcolsep}{4.2pt}

\begin{table}[b]
\begin{center}
\begin{tabular}{c||cccc||ccc}
\hline \hline
& & & & & & & \\[-2mm]
\etj ~range~ & $ ~d\sigma/d\etj$ & $\Delta_{stat.}$ & $\Delta_{syst.}$ & $\Delta_{e.s.}$ & $~(1+\delta_{frag.})$ & $(1+\delta_{u.e.})$ & $(1+\delta_{hadr.})$ \\[2mm]
$[\gev]$ & \multicolumn{4}{c||}{$[\pb/\gev]$} & & & \\[2mm]
\hline
\multicolumn{8}{c}{} \\[-2mm]
\multicolumn{1}{c}{} & \multicolumn{7}{c}{$95 \leq \Wgp \leq 285~\gev$} \\[2mm]
\hline
& & & & & & & \\[-3mm]
$21 \relbarx 28$&$65.4$ & {\footnotesize $\pm 0.6$} & {\footnotesize $\pm 1.6$}&$^{+5.5}_{-5.3}$&$0.95$ {\footnotesize $\pm 0.02$} & $1.08$ {\footnotesize $\pm 0.01$} & $1.03$ {\footnotesize $\pm 0.03$} \\[2mm]
$28 \relbarx 35$&$14.0$ & {\footnotesize $\pm 0.3$} & {\footnotesize $\pm 0.3$}&$^{+1.4}_{-1.2}$&$0.95$ {\footnotesize $\pm 0.02$} & $1.06$ {\footnotesize $\pm 0.02$} & $1.00$ {\footnotesize $\pm 0.04$} \\[2mm]
$35 \relbarx 42$&$3.56$ & {\footnotesize $\pm 0.14$} & {\footnotesize $\pm 0.09$}&$^{+0.39}_{-0.33}$&$0.95$ {\footnotesize $\pm 0.01$} & $1.04$ {\footnotesize $\pm 0.01$} & $1.00$ {\footnotesize $\pm 0.02$} \\[2mm]
$42 \relbarx 52$&$0.908$ & {\footnotesize $\pm 0.060$} & {\footnotesize $\pm 0.018$}&$^{+0.107}_{-0.095}$&$0.95$ {\footnotesize $\pm 0.01$} & $1.01$ {\footnotesize $\pm 0.02$} & $0.96$ {\footnotesize $\pm 0.01$} \\[2mm]
$52 \relbarx 62$&$0.192$ & {\footnotesize $\pm 0.028$} & {\footnotesize $\pm 0.012$}&$^{+0.023}_{-0.021}$&$0.97$ {\footnotesize $\pm 0.03$} & $1.01$ {\footnotesize $\pm 0.03$} & $0.98$ {\footnotesize $\pm 0.04$} \\[2mm]
$62 \relbarx 75$&$0.0483$ & {\footnotesize $\pm 0.0121$} & {\footnotesize $\pm 0.0018$}&$^{+0.0063}_{-0.0071}$&$0.96$ {\footnotesize $\pm 0.04$} & $1.00$ {\footnotesize $\pm 0.05$} & $0.96$ {\footnotesize $\pm 0.04$} \\[2mm] 
\hline
\multicolumn{8}{c}{} \\[-2mm]
\multicolumn{1}{c}{} & \multicolumn{7}{c}{$95 \leq \Wgp <  212~\gev$} \\[2mm]
\hline
& & & & & & & \\[-3mm]
$21 \relbarx 28$&$32.7$ & {\footnotesize $\pm 0.4$} & {\footnotesize $\pm 0.7$}&$^{+2.6}_{-3.0}$&$0.94$ {\footnotesize $\pm 0.02$} & $1.07$ {\footnotesize $\pm 0.01$} & $1.00$ {\footnotesize $\pm 0.03$} \\[2mm]
$28 \relbarx 35$&$6.21$ & {\footnotesize $\pm 0.19$} & {\footnotesize $\pm 0.18$}&$^{+0.60}_{-0.50}$&$0.94$ {\footnotesize $\pm 0.01$} & $1.04$ {\footnotesize $\pm 0.01$} & $0.98$ {\footnotesize $\pm 0.02$} \\[2mm]
$35 \relbarx 42$&$1.51$ & {\footnotesize $\pm 0.09$} & {\footnotesize $\pm 0.03$}&$^{+0.15}_{-0.14}$&$0.94$ {\footnotesize $\pm 0.02$} & $1.04$ {\footnotesize $\pm 0.01$} & $0.98$ {\footnotesize $\pm 0.01$} \\[2mm]
$42 \relbarx 52$&$0.236$ & {\footnotesize $\pm 0.030$} & {\footnotesize $\pm 0.008$}&$^{+0.025}_{-0.024}$&$0.93$ {\footnotesize $\pm 0.02$} & $1.00$ {\footnotesize $\pm 0.03$} & $0.93$ {\footnotesize $\pm 0.02$} \\[2mm]
$52 \relbarx 62$&$0.0360$ & {\footnotesize $\pm 0.0115$} & {\footnotesize $\pm 0.0009$}&$^{+0.0052}_{-0.0041}$&$0.92$ {\footnotesize $\pm 0.04$} & $1.00$ {\footnotesize $\pm 0.05$} & $0.92$ {\footnotesize $\pm 0.04$} \\[2mm]
$62 \relbarx 75$&$0.00511$ & {\footnotesize $\pm 0.00365$} & {\footnotesize $\pm 0.00019$}&$^{+0.00041}_{-0.00115}$&$0.90$ {\footnotesize $\pm 0.11$} & $0.98$ {\footnotesize $\pm 0.12$} & $0.88$ {\footnotesize $\pm 0.10$} \\[3mm]
\hline
\multicolumn{8}{c}{} \\[-2mm]
\multicolumn{1}{c}{} & \multicolumn{7}{c}{$212 \leq \Wgp \leq 285~\gev$} \\[2mm]
\hline
& & & & & & & \\[-3mm]
$21 \relbarx 28$&$32.8$ & {\footnotesize $\pm 0.4$} & {\footnotesize $\pm 1.0$}&$^{+2.8}_{-2.3}$&$0.97$ {\footnotesize $\pm 0.01$} & $1.10$ {\footnotesize $\pm 0.02$} & $1.06$ {\footnotesize $\pm 0.03$} \\[2mm]
$28 \relbarx 35$&$7.81$ & {\footnotesize $\pm 0.21$} & {\footnotesize $\pm 0.21$}&$^{+0.83}_{-0.71}$&$0.95$ {\footnotesize $\pm 0.02$} & $1.07$ {\footnotesize $\pm 0.03$} & $1.03$ {\footnotesize $\pm 0.06$} \\[2mm]
$35 \relbarx 42$&$2.05$ & {\footnotesize $\pm 0.11$} & {\footnotesize $\pm 0.07$}&$^{+0.23}_{-0.20}$&$0.96$ {\footnotesize $\pm 0.01$} & $1.05$ {\footnotesize $\pm 0.03$} & $1.01$ {\footnotesize $\pm 0.03$} \\[2mm]
$42 \relbarx 52$&$0.676$ & {\footnotesize $\pm 0.053$} & {\footnotesize $\pm 0.017$}&$^{+0.083}_{-0.071}$&$0.97$ {\footnotesize $\pm 0.02$} & $1.01$ {\footnotesize $\pm 0.02$} & $0.98$ {\footnotesize $\pm 0.02$} \\[2mm]
$52 \relbarx 62$&$0.157$ & {\footnotesize $\pm 0.026$} & {\footnotesize $\pm 0.012$}&$^{+0.018}_{-0.017}$&$0.99$ {\footnotesize $\pm 0.03$} & $1.01$ {\footnotesize $\pm 0.04$} & $1.00$ {\footnotesize $\pm 0.05$} \\[2mm]
$62 \relbarx 75$&$0.0434$ & {\footnotesize $\pm 0.0117$} & {\footnotesize $\pm 0.0018$}&$^{+0.0061}_{-0.0057}$&$0.97$ {\footnotesize $\pm 0.04$} & $1.00$ {\footnotesize $\pm 0.05$} & $0.97$ {\footnotesize $\pm 0.05$} \\[2mm] 
\hline \hline
\end{tabular}
\end{center}
\it \caption{\label{tab:fullinclhetet}Measured differential $e^+ p$ cross
section \setj ~for inclusive jet photoproduction ($Q^2\leq 1~\gevsq$),
integrated over $-1 \leq \etaj \leq 2.5$ in three regions of \Wgp. Jets are
defined using the inclusive $k_{\bot}$ algorithm. The statistical
($\Delta_{stat.}$),  systematic ($\Delta_{syst.}$) (excluding LAr energy scale)
and LAr energy scale ($\Delta_{e.s.}$) uncertainties are shown  separately. 
The correction factors applied to the NLO QCD predictions are also shown
separately as ($1+\delta_{frag.}$), for fragmentation, ($1+\delta_{u.e.}$), for
the underlying event, and the product ($1+\delta_{hadr.}$) for the total
hadronisation correction.} 
\end{table}

\clearpage

\renewcommand{\tabcolsep}{4.2pt}

\begin{table}[b]
\begin{center}
\begin{tabular}{c||cccc||ccc}
\hline \hline 
& & & & & & & \\[-2mm]
\etj ~range~ & $ ~d\sigma/d\etj$ & $\Delta_{stat.}$ & $\Delta_{syst.}$ & $\Delta_{e.s.}$ & $~(1+\delta_{frag.})$ & $(1+\delta_{u.e.})$ & $(1+\delta_{hadr.})$ \\[2mm]
$[\gev]$ & \multicolumn{4}{c||}{$[\pb/\gev]$} & & & \\[2mm]
\hline
\multicolumn{8}{c}{} \\[-2mm]
\multicolumn{1}{c}{} & \multicolumn{7}{c}{$164 \leq \Wgp \leq 242~\gev$ ~;~ $Q^2 \leq 0.01~\gevsq$} \\[2mm]
\hline
& & & & & & & \\[-3mm]
$ 5 \relbarx  8$ & $24600$ & {\footnotesize $\pm 200$} & {\footnotesize $\pm 1600$} & $ ^{+3000}_{-2900} $ & $ 0.72$ {\footnotesize $\pm 0.06$} & $1.77$ {\footnotesize $\pm 0.23$} & $1.25$ {\footnotesize $\pm 0.06$} \\[2mm]
$ 8 \relbarx 12$ & $3070$ & {\footnotesize $\pm 60$} & {\footnotesize $\pm 230$} & $ ^{+520}_{-470} $ & $ 0.80$ {\footnotesize $\pm 0.08$} & $1.66$ {\footnotesize $\pm 0.11$} & $1.31$ {\footnotesize $\pm 0.06$} \\[2mm]
$12 \relbarx 16$ & $505$ & {\footnotesize $\pm 26$} & {\footnotesize $\pm 30$} & $ ^{+94}_{-84} $ & $ 0.87$ {\footnotesize $\pm 0.08$} & $1.43$ {\footnotesize $\pm 0.10$} & $1.23$ {\footnotesize $\pm 0.04$} \\[2mm]
$16 \relbarx 21$ & $126$ & {\footnotesize $\pm 11$} & {\footnotesize $\pm 6$} & $ ^{+27}_{-19} $ & $ 0.83$ {\footnotesize $\pm 0.11$} & $1.26$ {\footnotesize $\pm 0.09$} & $1.04$ {\footnotesize $\pm 0.09$} \\[2mm]
$21 \relbarx 28$ & $28.3$ & {\footnotesize $\pm 6.1$} & {\footnotesize $\pm 4.6$} & $ ^{+10.1}_{-6.9} $ & $ 0.86$ {\footnotesize $\pm 0.17$} & $1.23$ {\footnotesize $\pm 0.13$} & $1.05$ {\footnotesize $\pm 0.17$} \\[2mm]              
\hline
\multicolumn{8}{c}{} \\[-2mm]
\multicolumn{1}{c}{} & \multicolumn{7}{c}{$164 \leq \Wgp \leq 242~\gev$ ~;~ $Q^2 \leq 1~\gevsq$} \\[2mm]
\hline
& & & & & & & \\[-3mm]
$21 \relbarx 28$ & $30.1$ & {\footnotesize $\pm 0.4$} & {\footnotesize $\pm 0.8$} & $ ^{+2.4}_{-2.5} $ & $ 0.95$ {\footnotesize $\pm 0.02$} & $1.09$ {\footnotesize $\pm 0.01$} & $1.04$ {\footnotesize $\pm 0.03$} \\[2mm]
$28 \relbarx 35$ & $6.74$ & {\footnotesize $\pm 0.19$} & {\footnotesize $\pm 0.18$} & $ ^{+0.63}_{-0.58} $ & $ 0.95$ {\footnotesize $\pm 0.01$} & $1.07$ {\footnotesize $\pm 0.02$} & $1.01$ {\footnotesize $\pm 0.04$} \\[2mm]
$35 \relbarx 42$ & $1.66$ & {\footnotesize $\pm 0.10$} & {\footnotesize $\pm 0.04$} & $ ^{+0.18}_{-0.14} $ & $ 0.96$ {\footnotesize $\pm 0.01$} & $1.04$ {\footnotesize $\pm 0.02$} & $1.00$ {\footnotesize $\pm 0.02$} \\[2mm]
$42 \relbarx 52$ & $0.417$ & {\footnotesize $\pm 0.041$} & {\footnotesize $\pm 0.013$} & $ ^{+0.043}_{-0.040} $ & $ 0.96$ {\footnotesize $\pm 0.02$} & $1.01$ {\footnotesize $\pm 0.03$} & $0.97$ {\footnotesize $\pm 0.02$} \\[2mm]
$52 \relbarx 62$ & $0.0773$ & {\footnotesize $\pm 0.0174$} & {\footnotesize $\pm 0.0066$} & $ ^{+0.0113}_{-0.0119} $ & $ 0.95$ {\footnotesize $\pm 0.03$} & $1.00$ {\footnotesize $\pm 0.04$} & $0.95$ {\footnotesize $\pm 0.03$} \\[2mm] 
$62 \relbarx 75$ & $0.0132$ & {\footnotesize $\pm 0.0059$} & {\footnotesize $\pm 0.0014$} & $ ^{+0.0016}_{-0.0019} $ & $ 0.94$ {\footnotesize $\pm 0.07$} & $0.95$ {\footnotesize $\pm 0.08$} & $0.89$ {\footnotesize $\pm 0.09$} \\[2mm] 
\hline \hline
\end{tabular}
\end{center}
\it \caption{\label{tab:combfullinclhetet}Measured differential $e^+ p$ cross
section \setj ~for inclusive jet photoproduction, integrated over $-1 \leq
\etaj \leq 2.5$ in the kinematic region $164 \leq \Wgp \leq 242~\gev$ (see Tab. 1 caption for further details).} 

\end{table}

\clearpage

\renewcommand{\tabcolsep}{4.2pt}

\begin{table}[t]
\begin{center}
\begin{tabular}{c||cccc||ccc}
\hline \hline
& & & & & & & \\[-3mm]
 \etaj ~range~ & $ ~d\sigma/d\etaj$ & $\Delta_{stat.}$ & $\Delta_{syst.}$ & $\Delta_{e.s.}$ & $(1+\delta_{frag.})$ & $(1+\delta_{u.e.})$ & $(1+\delta_{hadr.})$ \\[2mm]
 & \multicolumn{4}{c||}{$[\pb]$} & & & \\[2mm]
\hline
\multicolumn{8}{c}{} \\[-2mm]
\multicolumn{1}{c}{} & \multicolumn{7}{c}{$21 \leq \etj \leq 75~\gev$} \\[2mm]
\hline
& & & & & & & \\[-3mm]
$-1 \relbarx 0$  & $37.8$ & {\footnotesize $\pm 1.2$} & {\footnotesize $\pm 2.1$} & $ ^{+3.0}_{-2.5} $ & $ 0.84$ {\footnotesize $\pm 0.03$} & $1.04$ {\footnotesize $\pm 0.03$} & $0.87$ {\footnotesize $\pm 0.05$} \\[2mm]
$0 \relbarx 0.5$ & $173$ & {\footnotesize $\pm 4$} & {\footnotesize $\pm 4$} & $ ^{+12}_{-12} $ & $ 0.90$ {\footnotesize $\pm 0.03$} & $1.06$ {\footnotesize $\pm 0.02$} & $0.96$ {\footnotesize $\pm 0.05$} \\[2mm]
$0.5 \relbarx 1$ & $257$ & {\footnotesize $\pm 5$} & {\footnotesize $\pm 7$} & $ ^{+21}_{-23} $ & $ 0.94$ {\footnotesize $\pm 0.02$} & $1.06$ {\footnotesize $\pm 0.02$} & $1.00$ {\footnotesize $\pm 0.04$} \\[2mm]
$1 \relbarx 1.5$ & $253$ & {\footnotesize $\pm 4$} & {\footnotesize $\pm 6$} & $ ^{+25}_{-20} $ & $ 0.96$ {\footnotesize $\pm 0.01$} & $1.07$ {\footnotesize $\pm 0.02$} & $1.02$ {\footnotesize $\pm 0.03$} \\[2mm]
$1.5 \relbarx 2$ & $237$ & {\footnotesize $\pm 4$} & {\footnotesize $\pm 7$} & $ ^{+22}_{-19} $ & $ 0.99$ {\footnotesize $\pm 0.01$} & $1.09$ {\footnotesize $\pm 0.01$} & $1.08$ {\footnotesize $\pm 0.01$} \\[2mm]
$2 \relbarx 2.5$ & $186$ & {\footnotesize $\pm 4$} & {\footnotesize $\pm 4$} & $ ^{+17}_{-19} $ & $ 1.01$ {\footnotesize $\pm 0.01$} & $1.11$ {\footnotesize $\pm 0.02$} & $1.12$ {\footnotesize $\pm 0.01$} \\[2mm]
\hline
\multicolumn{8}{c}{} \\[-2mm]
\multicolumn{1}{c}{} & \multicolumn{7}{c}{$21 \leq \etj < 35~\gev$} \\[2mm]
\hline
& & & & & & & \\[-3mm]
$-1 \relbarx 0$  & $37.6$ & {\footnotesize $\pm 1.1$} & {\footnotesize $\pm 2.1$} & $ ^{+2.9}_{-2.5} $ & $ 0.84$ {\footnotesize $\pm 0.03$} & $1.04$ {\footnotesize $\pm 0.03$} & $0.87$ {\footnotesize $\pm 0.05$} \\[2mm]
$0 \relbarx 0.5$ & $166$ & {\footnotesize $\pm 4$} & {\footnotesize $\pm 4$} & $ ^{+11}_{-11} $ & $ 0.90$ {\footnotesize $\pm 0.03$} & $1.06$ {\footnotesize $\pm 0.02$} & $0.96$ {\footnotesize $\pm 0.05$} \\[2mm]
$0.5 \relbarx 1$ & $241$ & {\footnotesize $\pm 4$} & {\footnotesize $\pm 6$} & $ ^{+19}_{-21} $ & $ 0.94$ {\footnotesize $\pm 0.02$} & $1.07$ {\footnotesize $\pm 0.02$} & $1.00$ {\footnotesize $\pm 0.04$} \\[2mm]
$1 \relbarx 1.5$ & $233$ & {\footnotesize $\pm 4$} & {\footnotesize $\pm 6$} & $ ^{+23}_{-18} $ & $ 0.96$ {\footnotesize $\pm 0.02$} & $1.07$ {\footnotesize $\pm 0.02$} & $1.02$ {\footnotesize $\pm 0.03$} \\[2mm]
$1.5 \relbarx 2$ & $220$ & {\footnotesize $\pm 4$} & {\footnotesize $\pm 7$} & $ ^{+20}_{-18} $ & $ 0.99$ {\footnotesize $\pm 0.01$} & $1.10$ {\footnotesize $\pm 0.01$} & $1.09$ {\footnotesize $\pm 0.01$} \\[2mm]
$2 \relbarx 2.5$ & $174$ & {\footnotesize $\pm 4$} & {\footnotesize $\pm 4$} & $ ^{+16}_{-18} $ & $ 1.01$ {\footnotesize $\pm 0.01$} & $1.11$ {\footnotesize $\pm 0.02$} & $1.13$ {\footnotesize $\pm 0.02$} \\[2mm]
\hline
\multicolumn{8}{c}{} \\[-2mm]
\multicolumn{1}{c}{} & \multicolumn{7}{c}{$35 \leq \etj < 52~\gev$} \\[2mm]
\hline
& & & & & & & \\[-3mm]
$0 \relbarx 0.5$ & $7.56$ & {\footnotesize $\pm 0.75$} & {\footnotesize $\pm 0.64$} & $ ^{+1.08}_{-0.76} $ & $ 0.87$ {\footnotesize $\pm 0.03$} & $1.01$ {\footnotesize $\pm 0.02$} & $0.87$ {\footnotesize $\pm 0.03$} \\[2mm]
$0.5 \relbarx 1$ & $14.8$ & {\footnotesize $\pm 1.1$} & {\footnotesize $\pm 0.4$} & $ ^{+1.5}_{-1.4} $ & $ 0.94$ {\footnotesize $\pm 0.02$} & $1.03$ {\footnotesize $\pm 0.02$} & $0.96$ {\footnotesize $\pm 0.03$} \\[2mm]
$1 \relbarx 1.5$ & $18.3$ & {\footnotesize $\pm 1.2$} & {\footnotesize $\pm 0.4$} & $ ^{+1.8}_{-1.6} $ & $ 0.96$ {\footnotesize $\pm 0.01$} & $1.03$ {\footnotesize $\pm 0.02$} & $0.99$ {\footnotesize $\pm 0.02$} \\[2mm]
$1.5 \relbarx 2$ & $15.1$ & {\footnotesize $\pm 1.1$} & {\footnotesize $\pm 0.3$} & $ ^{+1.6}_{-1.4} $ & $ 0.98$ {\footnotesize $\pm 0.01$} & $1.04$ {\footnotesize $\pm 0.02$} & $1.02$ {\footnotesize $\pm 0.01$} \\[2mm]
$2 \relbarx 2.5$ & $11.5$ & {\footnotesize $\pm 1.0$} & {\footnotesize $\pm 0.2$} & $ ^{+1.5}_{-1.3} $ & $ 0.99$ {\footnotesize $\pm 0.02$} & $1.06$ {\footnotesize $\pm 0.02$} & $1.05$ {\footnotesize $\pm 0.02$} \\[2mm]
\hline
\multicolumn{8}{c}{} \\[-2mm]
\multicolumn{1}{c}{} & \multicolumn{7}{c}{$52 \leq \etj \leq 75~\gev$} \\[2mm]
\hline
& & & & & & & \\[-3mm]
$0.5 \relbarx 1$ & $1.16$ & {\footnotesize $\pm 0.33$} & {\footnotesize $\pm 0.08$} & $ ^{+0.08}_{-0.11} $ & $ 0.93$ {\footnotesize $\pm 0.05$} & $1.03$ {\footnotesize $\pm 0.05$} & $0.96$ {\footnotesize $\pm 0.05$} \\[2mm]
$1 \relbarx 1.5$ & $1.69$ & {\footnotesize $\pm 0.37$} & {\footnotesize $\pm 0.10$} & $ ^{+0.21}_{-0.23} $ & $ 0.97$ {\footnotesize $\pm 0.04$} & $0.97$ {\footnotesize $\pm 0.05$} & $0.94$ {\footnotesize $\pm 0.06$} \\[2mm]
$1.5 \relbarx 2$ & $1.84$ & {\footnotesize $\pm 0.39$} & {\footnotesize $\pm 0.12$} & $ ^{+0.25}_{-0.19} $ & $ 0.99$ {\footnotesize $\pm 0.04$} & $1.03$ {\footnotesize $\pm 0.04$} & $1.02$ {\footnotesize $\pm 0.04$} \\[2mm]
$2 \relbarx 2.5$ & $0.458$ & {\footnotesize $\pm 0.189$} & {\footnotesize $\pm 0.040$} & $ ^{+0.077}_{-0.069} $ & $ 0.98$ {\footnotesize $\pm 0.06$} & $1.04$ {\footnotesize $\pm 0.07$} & $1.02$ {\footnotesize $\pm 0.08$} \\[2mm]        
\hline \hline
\end{tabular}
\end{center}
\it \caption{\label{tab:fullinclhetrp}Measured differential $e^+ p$ cross
section \setaj ~for inclusive jet photoproduction ($Q^2\leq 1~\gevsq$),
integrated over four \etj ~ranges in the kinematic region $95 \leq \Wgp \leq
285~\gev$ (see Tab. 1 caption for further details).} 

\end{table}

\clearpage 

\renewcommand{\tabcolsep}{4.2pt}

\begin{table}[b]
\begin{center}
\begin{tabular}{c||cccc||ccc}
\hline \hline
& & & & & & & \\[-3mm]
 \etaj ~range~ & $ ~d\sigma/d\etaj$ & $\Delta_{stat.}$ & $\Delta_{syst.}$ & $\Delta_{e.s.}$ & $(1+\delta_{frag.})$ & $(1+\delta_{u.e.})$ & $(1+\delta_{hadr.})$ \\
& \multicolumn{4}{c||}{$[\pb]$} & & & \\[1.5mm]
\hline
\multicolumn{8}{c}{} \\[-2.5mm]
\multicolumn{1}{c}{} & \multicolumn{7}{c}{$21 \leq \etj < 35~\gev$ ~ ; ~ $95 \leq \Wgp <  212~\gev$} \\[1.5mm]
\hline
& & & & & & & \\[-3.5mm]
$0 \relbarx 0.5$ & $32.6$ & {\footnotesize $\pm 1.5$} & {\footnotesize $\pm 0.9$} & $ ^{+1.7}_{-3.2} $ & $ 0.79$ {\footnotesize $\pm 0.04$} & $1.02$ {\footnotesize $\pm 0.04$} & $0.81$ {\footnotesize $\pm 0.06$} \\[1mm]
$0.5 \relbarx 1$ & $114$ & {\footnotesize $\pm 3$} & {\footnotesize $\pm 4$} & $ ^{+10}_{-9} $ & $ 0.90$ {\footnotesize $\pm 0.02$} & $1.06$ {\footnotesize $\pm 0.02$} & $0.95$ {\footnotesize $\pm 0.03$} \\[1mm]
$1 \relbarx 1.5$ & $141$ & {\footnotesize $\pm 3$} & {\footnotesize $\pm 3$} & $ ^{+12}_{-12} $ & $ 0.93$ {\footnotesize $\pm 0.02$} & $1.05$ {\footnotesize $\pm 0.02$} & $0.98$ {\footnotesize $\pm 0.04$} \\[1mm]
$1.5 \relbarx 2$ & $142$ & {\footnotesize $\pm 3$} & {\footnotesize $\pm 5$} & $ ^{+13}_{-11} $ & $ 0.98$ {\footnotesize $\pm 0.01$} & $1.06$ {\footnotesize $\pm 0.02$} & $1.04$ {\footnotesize $\pm 0.02$} \\[1mm]
$2 \relbarx 2.5$ & $114$ & {\footnotesize $\pm 3$} & {\footnotesize $\pm 3$} & $ ^{+9}_{-14} $ & $ 0.99$ {\footnotesize $\pm 0.02$} & $1.10$ {\footnotesize $\pm 0.03$} & $1.09$ {\footnotesize $\pm 0.02$} \\[1.5mm]
\hline
\multicolumn{8}{c}{} \\[-2.5mm]
\multicolumn{1}{c}{} & \multicolumn{7}{c}{$21 \leq \etj < 35~\gev$ ~ ; ~ $212 \leq \Wgp \leq 285~\gev$} \\[1.5mm]
\hline
& & & & & & & \\[-3.5mm]
$-1 \relbarx 0$  & $37.3$ & {\footnotesize $\pm 1.2$} & {\footnotesize $\pm 1.9$} & $ ^{+3.1}_{-2.9} $ & $ 0.85$ {\footnotesize $\pm 0.03$} & $1.04$ {\footnotesize $\pm 0.03$} & $0.88$ {\footnotesize $\pm 0.05$} \\[1mm]
$0 \relbarx 0.5$ & $133$ & {\footnotesize $\pm 3$} & {\footnotesize $\pm 4$} & $ ^{+9}_{-8} $ & $ 0.94$ {\footnotesize $\pm 0.03$} & $1.07$ {\footnotesize $\pm 0.02$} & $1.01$ {\footnotesize $\pm 0.05$} \\[1mm]
$0.5 \relbarx 1$ & $127$ & {\footnotesize $\pm 3$} & {\footnotesize $\pm 6$} & $ ^{+9}_{-12} $ & $ 0.97$ {\footnotesize $\pm 0.02$} & $1.08$ {\footnotesize $\pm 0.04$} & $1.05$ {\footnotesize $\pm 0.05$} \\[1mm]
$1 \relbarx 1.5$ & $91.7$ & {\footnotesize $\pm 2.6$} & {\footnotesize $\pm 3.7$} & $ ^{+11.6}_{-6.2} $ & $ 1.01$ {\footnotesize $\pm 0.02$} & $1.10$ {\footnotesize $\pm 0.03$} & $1.11$ {\footnotesize $\pm 0.03$} \\[1mm]
$1.5 \relbarx 2$ & $78.7$ & {\footnotesize $\pm 2.6$} & {\footnotesize $\pm 2.2$} & $ ^{+6.8}_{-6.3} $ & $ 1.02$ {\footnotesize $\pm 0.02$} & $1.16$ {\footnotesize $\pm 0.03$} & $1.19$ {\footnotesize $\pm 0.03$} \\[1mm]
$2 \relbarx 2.5$ & $59.6$ & {\footnotesize $\pm 2.3$} & {\footnotesize $\pm 2.4$} & $ ^{+7.4}_{-3.9} $ & $ 1.05$ {\footnotesize $\pm 0.03$} & $1.14$ {\footnotesize $\pm 0.03$} & $1.20$ {\footnotesize $\pm 0.03$} \\[1.5mm]
\hline
\multicolumn{8}{c}{} \\[-2.5mm]
\multicolumn{1}{c}{} & \multicolumn{7}{c}{$35 \leq \etj < 52~\gev$ ~ ; ~ $95 \leq \Wgp <  212~\gev$} \\[1.5mm]
\hline
& & & & & & & \\[-3.5mm]
$0.5 \relbarx 1$ & $2.72$ & {\footnotesize $\pm 0.46$} & {\footnotesize $\pm 0.17$} & $ ^{+0.16}_{-0.18} $ & $ 0.82$ {\footnotesize $\pm 0.04$} & $1.03$ {\footnotesize $\pm 0.04$} & $0.85$ {\footnotesize $\pm 0.05$} \\[1mm]
$1 \relbarx 1.5$ & $8.07$ & {\footnotesize $\pm 0.81$} & {\footnotesize $\pm 0.19$} & $ ^{+0.76}_{-0.72} $ & $ 0.93$ {\footnotesize $\pm 0.02$} & $1.02$ {\footnotesize $\pm 0.02$} & $0.95$ {\footnotesize $\pm 0.02$} \\[1mm]
$1.5 \relbarx 2$ & $8.27$ & {\footnotesize $\pm 0.81$} & {\footnotesize $\pm 0.20$} & $ ^{+0.87}_{-0.75} $ & $ 0.96$ {\footnotesize $\pm 0.02$} & $1.02$ {\footnotesize $\pm 0.02$} & $0.98$ {\footnotesize $\pm 0.02$} \\[1mm]
$2 \relbarx 2.5$ & $6.57$ & {\footnotesize $\pm 0.71$} & {\footnotesize $\pm 0.19$} & $ ^{+0.84}_{-0.71} $ & $ 0.97$ {\footnotesize $\pm 0.03$} & $1.05$ {\footnotesize $\pm 0.03$} & $1.02$ {\footnotesize $\pm 0.02$} \\[1.5mm]
\hline
\multicolumn{8}{c}{} \\[-2.5mm]
\multicolumn{1}{c}{} & \multicolumn{7}{c}{$35 \leq \etj < 52~\gev$ ~ ; ~ $212 \leq \Wgp \leq 285~\gev$} \\[1.5mm]
\hline
& & & & & & & \\[-3.5mm]
$0 \relbarx 0.5$ & $7.44$ & {\footnotesize $\pm 0.75$} & {\footnotesize $\pm 0.62$} & $ ^{+1.04}_{-0.72} $ & $ 0.87$ {\footnotesize $\pm 0.03$} & $1.01$ {\footnotesize $\pm 0.02$} & $0.88$ {\footnotesize $\pm 0.03$} \\[1mm]
$0.5 \relbarx 1$ & $12.1$ & {\footnotesize $\pm 1.0$} & {\footnotesize $\pm 0.4$} & $ ^{+1.3}_{-1.2}   $ & $ 0.96$ {\footnotesize $\pm 0.02$} & $1.03$ {\footnotesize $\pm 0.02$} & $0.99$ {\footnotesize $\pm 0.03$} \\[1mm]
$1 \relbarx 1.5$ & $10.2$ & {\footnotesize $\pm 0.9$} & {\footnotesize $\pm 0.2$} & $ ^{+1.0}_{-0.9}   $ & $ 0.99$ {\footnotesize $\pm 0.02$} & $1.04$ {\footnotesize $\pm 0.02$} & $1.02$ {\footnotesize $\pm 0.03$} \\[1mm]
$1.5 \relbarx 2$ & $6.85$ & {\footnotesize $\pm 0.77$} & {\footnotesize $\pm 0.19$} & $ ^{+0.77}_{-0.71} $ & $ 1.00$ {\footnotesize $\pm 0.02$} & $1.07$ {\footnotesize $\pm 0.02$} & $1.08$ {\footnotesize $\pm 0.02$} \\[1mm]
$2 \relbarx 2.5$ & $4.97$ & {\footnotesize $\pm 0.63$} & {\footnotesize $\pm 0.13$} & $ ^{+0.62}_{-0.56} $ & $ 1.02$ {\footnotesize $\pm 0.03$} & $1.07$ {\footnotesize $\pm 0.04$} & $1.09$ {\footnotesize $\pm 0.03$} \\[1.5mm]
\hline
\multicolumn{8}{c}{} \\[-2.5mm]
\multicolumn{1}{c}{} & \multicolumn{7}{c}{$52 \leq \etj \leq 75~\gev$ ~ ; ~ $212 \leq \Wgp \leq 285~\gev$} \\[1.5mm]
\hline
& & & & & & & \\[-3.5mm]
$0.5 \relbarx 1$ & $1.17$ & {\footnotesize $\pm 0.33$} & {\footnotesize $\pm 0.08$} & $ ^{+0.07}_{-0.11} $ & $ 0.94$ {\footnotesize $\pm 0.05$} & $1.02$ {\footnotesize $\pm 0.05$} & $0.96$ {\footnotesize $\pm 0.05$} \\[1mm]
$1 \relbarx 1.5$ & $1.49$ & {\footnotesize $\pm 0.35$} & {\footnotesize $\pm 0.08$} & $ ^{+0.17}_{-0.18} $ & $ 0.99$ {\footnotesize $\pm 0.04$} & $0.96$ {\footnotesize $\pm 0.06$} & $0.96$ {\footnotesize $\pm 0.07$} \\[1mm]
$1.5 \relbarx 2$ & $1.24$ & {\footnotesize $\pm 0.32$} & {\footnotesize $\pm 0.12$} & $ ^{+0.19}_{-0.15} $ & $ 1.03$ {\footnotesize $\pm 0.06$} & $1.02$ {\footnotesize $\pm 0.05$} & $1.05$ {\footnotesize $\pm 0.07$} \\[1mm]
$2 \relbarx 2.5$ & $0.423$ & {\footnotesize $\pm 0.191$} & {\footnotesize $\pm 0.045$} & $ ^{+0.073}_{-0.058} $ & $ 1.02$ {\footnotesize $\pm 0.08$} & $1.15$ {\footnotesize $\pm 0.18$} & $1.17$ {\footnotesize $\pm 0.18$} \\[1.5mm]        
\hline \hline
\end{tabular}
\end{center}
\it \caption{\label{tab:2y3etinclheteta}Measured differential $e^+ p$ cross
section \setaj ~for inclusive jet photoproduction  ($Q^2 \leq 1~\gevsq$). The phase space of the measurement is divided into two regions of \Wgp ~and three regions of \etj ~(see Tab. 1 caption for further details).} 

\end{table}

\clearpage 

\renewcommand{\tabcolsep}{4.2pt}

\begin{table}[t]
\begin{center}
\begin{tabular}{c||cccc||ccc}
\hline \hline
& & & & & & & \\[-3mm]
 \etaj ~range~& $ ~d\sigma/d\etaj$ & $\Delta_{stat.}$ & $\Delta_{syst.}$ & $\Delta_{e.s.}$ & $(1+\delta_{frag.})$ & $(1+\delta_{u.e.})$ & $(1+\delta_{hadr.})$ \\[2mm]
& \multicolumn{4}{c||}{$[\nb]$} & & & \\[2mm]
\hline
\multicolumn{8}{c}{} \\[-2mm]
\multicolumn{1}{c}{} & \multicolumn{7}{c}{$5 \leq \etj < 12~\gev$} \\[2mm]
\hline
& & & & & & & \\[-3mm]
$-1 \relbarx -0.5$ & $16.2$ & {\footnotesize $\pm 0.4$} & {\footnotesize $\pm 1.1$} & $ ^{+2.0}_{-1.8} $ & $ 0.79$ {\footnotesize $\pm 0.06$} & $1.29$ {\footnotesize $\pm 0.06$} & $1.02$ {\footnotesize $\pm 0.03$} \\[2mm]
$-0.5 \relbarx 0$ & $18.1$ & {\footnotesize $\pm 0.4$} & {\footnotesize $\pm 1.4$} & $ ^{+2.1}_{-2.0} $ & $ 0.80$ {\footnotesize $\pm 0.06$} & $1.45$ {\footnotesize $\pm 0.08$} & $1.15$ {\footnotesize $\pm 0.02$} \\[2mm]
$0 \relbarx 0.5$ & $20.6$ & {\footnotesize $\pm 0.4$} & {\footnotesize $\pm 1.4$} & $ ^{+2.5}_{-2.3} $ & $ 0.73$ {\footnotesize $\pm 0.06$} & $1.64$ {\footnotesize $\pm 0.13$} & $1.19$ {\footnotesize $\pm 0.02$} \\[2mm]
$0.5 \relbarx 1$ & $24.1$ & {\footnotesize $\pm 0.5$} & {\footnotesize $\pm 1.7$} & $ ^{+2.7}_{-2.8} $ & $ 0.68$ {\footnotesize $\pm 0.06$} & $1.89$ {\footnotesize $\pm 0.27$} & $1.27$ {\footnotesize $\pm 0.06$} \\[2mm]
$1 \relbarx 1.5$ & $25.8$ & {\footnotesize $\pm 0.4$} & {\footnotesize $\pm 1.7$} & $ ^{+3.1}_{-2.8} $ & $ 0.69$ {\footnotesize $\pm 0.06$} & $1.99$ {\footnotesize $\pm 0.32$} & $1.35$ {\footnotesize $\pm 0.10$} \\[2mm]
\hline
\multicolumn{8}{c}{} \\[-2mm]
\multicolumn{1}{c}{} & \multicolumn{7}{c}{$12 \leq \etj \leq 21~\gev$} \\[2mm]
\hline
& & & & & & & \\[-3mm]
$-0.5 \relbarx 0$ & $0.584$ & {\footnotesize $\pm 0.081$} & {\footnotesize $\pm 0.034$} & $ ^{+0.063}_{-0.102} $ & $ 0.77$ {\footnotesize $\pm 0.11$} & $1.12$ {\footnotesize $\pm 0.07$} & $0.87$ {\footnotesize $\pm 0.11$} \\[2mm]        
$0 \relbarx 0.5$ & $0.987$ & {\footnotesize $\pm 0.110$} & {\footnotesize $\pm 0.080$} & $ ^{+0.106}_{-0.102} $ & $ 0.91$ {\footnotesize $\pm 0.09$} & $1.22$ {\footnotesize $\pm 0.06$} & $1.11$ {\footnotesize $\pm 0.11$} \\[2mm]        
$0.5 \relbarx 1$ & $1.02$ & {\footnotesize $\pm 0.11$} & {\footnotesize $\pm 0.07$} & $ ^{+0.09}_{-0.20} $ & $ 0.89$ {\footnotesize $\pm 0.09$} & $1.33$ {\footnotesize $\pm 0.08$} & $1.19$ {\footnotesize $\pm 0.09$} \\[2mm]
$1 \relbarx 1.5$ & $0.803$ & {\footnotesize $\pm 0.096$} & {\footnotesize $\pm 0.060$} & $ ^{+0.175}_{-0.110} $ & $ 0.91$ {\footnotesize $\pm 0.07$} & $1.53$ {\footnotesize $\pm 0.14$} & $1.39$ {\footnotesize $\pm 0.08$} \\[2mm]        
$1.5 \relbarx 2$ & $1.12$ & {\footnotesize $\pm 0.11$} & {\footnotesize $\pm 0.08$} & $ ^{+0.26}_{-0.15} $ & $ 0.83$ {\footnotesize $\pm 0.14$} & $1.74$ {\footnotesize $\pm 0.33$} & $1.41$ {\footnotesize $\pm 0.08$} \\[2mm]
$2 \relbarx 2.5$ & $0.824$ & {\footnotesize $\pm 0.078$} & {\footnotesize $\pm 0.056$} & $ ^{+0.264}_{-0.171} $ & $ 0.88$ {\footnotesize $\pm 0.09$} & $1.75$ {\footnotesize $\pm 0.27$} & $1.53$ {\footnotesize $\pm 0.14$} \\[2mm]        
\hline \hline
\end{tabular}
\end{center}
\it \caption{\label{tab:fullinclhetrplow}Measured differential $e^+ p$ cross
section \setaj ~for inclusive jet photoproduction ($Q^2 \leq 0.01~\gevsq$),
integrated over two \etj ~ranges in the kinematic region $164 \leq \Wgp \leq
242~\gev$ (see Tab. 1 caption for further details).} 

\end{table}

\clearpage 

\renewcommand{\tabcolsep}{4.2pt}

\begin{table}[b]
\begin{center}
\begin{tabular}{cc||cccc}
\hline \hline 
& & & & & \\[-3mm]
\etj ~range~ & $\langle \etj \rangle$~ & $ ~d\sigma/d\etj$ & $\Delta_{stat.}$ & $\Delta_{syst.}$ & $\Delta_{e.s.}$ \\[2mm]
\multicolumn{2}{c||}{$[\gev]$} & \multicolumn{4}{c}{$[\pb/\gev]$} \\[2mm]
\hline
\multicolumn{6}{c}{} \\[-3mm]
\multicolumn{6}{c}{$164 \leq \Wgp \leq 242~\gev$ ~;~ $Q^2 \leq 0.01~\gevsq$} \\[2mm]
\hline
& & & & & \\[-3mm]
$ 5 \relbarx  8$ & $6.1$ & $39900$ & {\footnotesize $\pm 200$} & {\footnotesize $\pm 2600$} & $ ^{+4700}_{-4400} $ \\[2mm]
$ 8 \relbarx 12$ & $9.3$ & $3840$ & {\footnotesize $\pm 60$} & {\footnotesize $\pm 250$} & $ ^{+690}_{-570} $ \\[2mm]
$12 \relbarx 16$ & $14.3$ & $595$ & {\footnotesize $\pm 28$} & {\footnotesize $\pm 35$} & $ ^{+91}_{-82} $ \\[2mm]
$16 \relbarx 21$ & $18.1$ & $126$ & {\footnotesize $\pm 12$} & {\footnotesize $\pm 7$} & $ ^{+33}_{-18} $ \\[2mm]
\hline
\multicolumn{6}{c}{} \\[-3mm]
\multicolumn{6}{c}{$164 \leq \Wgp \leq 242~\gev$ ~;~ $Q^2 \leq 1~\gevsq$} \\[2mm]
\hline
& & & & & \\[-3mm]
$21 \relbarx 28$ & $23.6$ & $31.3$ & {\footnotesize $\pm 0.4$} & {\footnotesize $\pm 0.8$} & $ ^{+3.0}_{-2.9} $ \\[2mm]
$28 \relbarx 35$ & $30.8$ & $6.66$ & {\footnotesize $\pm 0.19$} & {\footnotesize $\pm 0.18$} & $ ^{+0.56}_{-0.57} $ \\[2mm]
$35 \relbarx 42$ & $37.8$ & $1.73$ & {\footnotesize $\pm 0.10$} & {\footnotesize $\pm 0.06$} & $ ^{+0.19}_{-0.16} $ \\[2mm]
$42 \relbarx 52$ & $45.8$ & $0.415$ & {\footnotesize $\pm 0.042$} & {\footnotesize $\pm 0.014$} & $ ^{+0.047}_{-0.044} $ \\[2mm]
$52 \relbarx 62$ & $55.8$ & $0.0794$ & {\footnotesize $\pm 0.0179$} & {\footnotesize $\pm 0.0048$} & $ ^{+0.0131}_{-0.0100} $ \\[2mm] 
$62 \relbarx 75$ & $66.4$ & $0.0143$ & {\footnotesize $\pm 0.0065$} & {\footnotesize $\pm 0.0004$} & $ ^{+0.0023}_{-0.0021} $ \\[2mm]   
\hline \hline
\end{tabular}
\end{center}
\it \caption{\label{tab:combfullinclhetetco}Measured differential $e^+ p$ cross
section \setj ~for inclusive jet photoproduction integrated over $-1 \leq \etaj
\leq 2.5$ in the kinematic region $164 \leq \Wgp \leq 242~\gev$.  Jets are
defined using the cone algorithm with $R=1$. For each range of \etj, the
average value of \etj calculated with PYTHIA is given in the second column. The
statistical ($\Delta_{stat.}$),  systematic ($\Delta_{syst.}$) and LAr energy
scale ($\Delta_{e.s.}$) uncertainties are shown  separately.} 
\end{table}

\renewcommand{\tabcolsep}{8pt}

\begin{table}[b]
\begin{center}
\begin{tabular}{c||cccc}
\hline \hline 
& & & & \\[-3mm]
$x_T$ & $S(x_T)$ & $\Delta_{stat.}$ & $\Delta_{syst.}$ & $\Delta_{e.s.}$  \\[2mm]
\hline
& & & & \\[-3mm]
$0.09$ & $0.0441$ & {\footnotesize $\pm 0.0011$} & {\footnotesize $\pm 0.0024$} & {\footnotesize $\pm 0.0078$} \\[2mm]
$0.13$ & $0.0201$ & {\footnotesize $\pm 0.0014$} & {\footnotesize $\pm 0.0012$} & {\footnotesize $\pm 0.0028$} \\[2mm]
$0.18$ & $0.00658$ & {\footnotesize $\pm 0.00106$} & {\footnotesize $\pm 0.00068$} & {\footnotesize $\pm 0.00241$} \\[2mm]
$0.25$ & $0.00236$ & {\footnotesize $\pm 0.00005$} & {\footnotesize $\pm 0.00015$} & {\footnotesize $\pm 0.00022$} \\[2mm]
$0.40$ & $0.000684$ & {\footnotesize $\pm 0.000053$} & {\footnotesize $\pm 0.000016$} & {\footnotesize $\pm 0.000073$} \\[2mm]
$0.56$ & $0.000185$ & {\footnotesize $\pm 0.000047$} & {\footnotesize $\pm 0.000009$} & {\footnotesize $\pm 0.000028$} \\[2mm]
\hline \hline
\end{tabular}
\end{center}
\it \caption{\label{tab:combxt} Scaled $\gamma p$ cross section at $\Wgp = 200
~\gev$  
as a function of $x_T$ for $|\eta^\star| \leq 0.5$. Jets are found
with the cone algorithm ($R=1$). The statistical ($\Delta_{stat.}$), 
systematic ($\Delta_{syst.}$) and LAr energy scale ($\Delta_{e.s.}$)
uncertainties are shown  separately.} 
\end{table}

\clearpage 

\clearpage 

\clearpage 

\clearpage 

\clearpage 

\clearpage 

\clearpage 

\clearpage 

%\begin{figure}[b]
%\begin{center}
%\vspace*{-0.3cm}
%\includegraphics[width=\textwidth]{jetp_9p.eps}
%\end{center}
%\vspace*{-0.3cm}
%\it \caption{\label{fig:fxT}Scaled $\gamma p$ cross section at 
%$\Wgp = 200 ~\gev$ 
%for inclusive jet production as a function of $x_T$ for 
%$|\eta^\star| \leq 0.5$. Jets are found with the cone algorithm ($R=1$).
%The data are compared with measurements
%from UA1~\cite{Arnison:1986vk,Albajar:1988tt} and D0~\cite{Abbott:2001ce,Abbott:1998ya} 
% of inclusive jet production  in $p\overline{p}$
%collisions at various cms energies.
%The predictions of PYTHIA for $\gamma p$, and for the resolved photon contribution are also shown multiplied by a factor 1.2.
%}
%\end{figure}

\end{document}